\documentclass[useAMS,usenatbib]{mn2e}

\usepackage{graphicx}
\usepackage{amstext}
\usepackage{amsmath}
\usepackage{amssymb}
\usepackage{color}
\usepackage[normalem]{ulem}
\usepackage{multirow}
\usepackage{longtable}
\usepackage{bm}
\usepackage{rotating}
\usepackage{pifont}

\newcommand{\cmark}{\color{blue}{\ding{51}}}%
\newcommand{\xmark}{\color{red}{\ding{55}}}%

\makeatletter
\def\@cline#1-#2\@nil{%
  \omit
  \@multicnt#1%
  \advance\@multispan\m@ne
  \ifnum\@multicnt=\@ne\@firstofone{&\omit}\fi
  \@multicnt#2%
  \advance\@multicnt-#1%
  \advance\@multispan\@ne
  \leaders\hrule\@height\arrayrulewidth\hfill
  \cr
  \noalign{\nobreak\vskip-\arrayrulewidth}}
\makeatother

\begin{document}

\title[Thermonuclear detonations ensuing WD mergers]{Thermonuclear
  detonations ensuing white dwarf mergers} 

\author[M. Dan, J. Guillochon, M. Br\" uggen, E. Ramirez-Ruiz,
S. Rosswog]{M. Dan$^{1}$\thanks{E-mail:
    marius.dan@hs.uni-hamburg.de}, J. Guillochon$^{2,3}$, M.
  Br\" uggen$^{1}$, E. Ramirez-Ruiz$^{4,5}$, S. Rosswog$^{6}$\\
$^{1}${Hamburger Sternwarte, Universit\"at Hamburg,
  Gojenbergsweg 112, 21029 Hamburg, Germany}\\
$^{2}${Harvard-Smithsonian Center for Astrophysics, The Institute for Theory and
Computation, 60 Garden Street, Cambridge, MA 02138, USA}\\
$^{3}${Einstein Fellow}\\
$^{4}${Department of Astronomy and Astrophysics, University of California, Santa Cruz, CA 95064, USA}\\
$^{5}${Radcliffe Fellow}\\
$^{6}${Astronomy and Oskar Klein Centre, Stockholm University, AlbaNova,
SE-106 91 Stockholm, Sweden}
}

\date{Accepted ?. Received ?; in original form ?}

\pagerange{\pageref{firstpage}--\pageref{lastpage}} \pubyear{2014}

\maketitle

\label{firstpage}

\def\msun{$M_{\odot}$}
\def\Msun{$M_{\odot}$ }

\begin{abstract}
The merger of two white dwarfs (WDs) has for many years not been considered as
the favoured model for the progenitor system of type Ia supernovae
(SNe Ia). But recent years have seen a change of opinion as a number of studies,
both observational and theoretical, have concluded that they should
contribute significantly to the observed type Ia supernova rate.
In this paper, we study the ignition and propagation of detonation through
post-merger remnants and we follow the resulting nucleosynthesis up to the
point where a homologous expansion is reached. 
In our study we cover the entire range of WD masses and
compositions. 
For the emergence of a detonation we study three different setups. 
The first two are guided by the merger remnants from our earlier 
simulations \citep{dan14}, while for the third one the ignitions were 
set by placing hotspots with properties determined by spatially 
resolved calculations taken from the literature.
There are some caveats to our approach which we investigate.
We carefully compare the nucleosynthetic
yields of successful explosions with SN Ia observations. Only three of our models
are consistent with all the imposed constraints and potentially lead
to a standard type Ia event. The first one, a $0.45\,M_\odot\ {\rm helium\
  (He)} + 0.9 \,M_\odot$ carbon-oxygen (CO) WD system produces a
sub-luminous, SN 1991bg-like event while the other two, a
$0.45\,M_\odot\ {\rm He} + 1.1\,M_\odot$ oxygen-neon (ONe) WD system and a
$1.05+1.05\,M_\odot$ system with two CO WDs, are good candidates for
common SNe Ia.  
\end{abstract}


\section{Introduction}

Thermonuclear supernovae have long been used as cosmological distance
indicators. By comparing brightness and redshifts of distant and nearby SNe Ia, it
appears that the rate of expansion of the universe is 
increasing \citep{riess98,perlmutter98}. 
However, their stellar progenitors have so far remained elusive.
There is a consensus that the exploding star is a WD made primarily of  CO that
has a binary companion which donates mass, 
but the identity of the companion and the exact nature of the
explosion mechanism have remained unclear \citep{maoz14}. 
Traditionally, three main scenarios have been proposed to explain the
bulk of SNe Ia \citep[for reviews,
see][]{hillebrandt00,howell11,wang12,postnov14,maoz14}. In the so-called
``single degenerate'' 
scenario, the CO WD accretes matter from a main
sequence, a sub-giant or a red-giant star up to near
the Chandrasekhar mass limit, when carbon ignition near the center,
followed by a deflagration and later a transition to a detonation,
produces a SN Ia-like event \citep{whelan73,nomoto82a}. 
In the ``sub-Chandrasekhar''  scenario, the CO WD is
ignited before it reaches the Chandrasekhar mass 
\citep{nomoto82b,livne90,woosley94,livne95,garcia99,fink07,guillochon10,sim10,woosley11,sim12,dan12,moll13}. 
In this scenario, a first detonation occurs in the accreted helium
material. The resulting shock waves are propagating 
through the core and may trigger a second detonation close to the
interface between the core and the envelope (also known as “edge-lit
detonation” model) or close to the core's centre, where the 
compressional waves converge \citep{seitenzahl09,shen14a}. 
The second detonation would disrupt the central remnant completely
and produce a SN Ia-like event. If only the He shell detonates this
could produce a roughly ten times fainter 
SN Ia, sometimes referred to as SNe ``point'' Ia
\citep{bildsten07,shen09,waldman11,holcomb13}.  
The ``double degenerate'' model involves two CO WDs in a close binary system
being drawn together as they lose angular momentum by radiating
gravitational waves that eventually merges \citep{iben84,webbink84}. A
thermonuclear explosion could be ignited 
at different stages of the WD-WD binary evolution, either during mass
transfer \citep{guillochon10,pakmor13}, during the merger
\citep{pakmor10,raskin12,pakmor12,dan12,moll13,sato15} or after the
merger, in the remnant phase \citep{dan14,raskin14,kashyap15,sato15}. 
Another scenario that could contribute by more than 20\%
\citep{tsebrenko15} to the SN Ia rate is the ``core-degenerate''
scenario where the WD merges with the core of an asymptotic giant
branch star \citep{livio03,kashi11}.
These scenarios are meant to explain the bulk of SNe Ia.
They may be complemented by rarer channels that trigger thermonuclear
explosions in WDs that occur under less common initial conditions.
For example, also the compression by the tidal field of a moderately massive
black hole may cause a type Ia-like thermonuclear explosion
\citep{luminet89,rosswog09b}, or, collisions of WDs as they occur 
in globular cluster cores can produce strong thermonuclear explosions
\citep{rosswog09c,raskin09,kushnir13}. While the former would 
typically cause very peculiar events, the latter case produces for the most common
WDs around $0.6\,M_\odot$ explosions signatures that are similar to typical
type Ia events \citep{rosswog09c}. Both types of events, however,
are most likely too rare to explain the bulk of type Ia events.

The present study will focus on the post-merger remnant phase. The
merger remnants show a similar thermodynamic and rotational structure:
a cold, isothermal core surrounded by a hot, 
pressure supported envelope, a rotationally supported disk and a tidal
tail \citep[e.g.,][]{guerrero04,yoon07,zhu13,dan14}.
It is inside the hot envelope region of the remnant where the 
nuclear combustible (helium and, maybe, carbon/oxygen for the most
massive WD mergers) may undergo a thermonuclear runaway and a
detonation.  From the moment when a detonation is ignited, the
evolution is very similar to the sub-Chandrasekhar model.  

In \cite{dan14} we had investigated possible detonations from WD
merger remnants by comparing local dynamical and nuclear burning time
scales. Dynamical burning, however, is a necessary but not sufficient
criterion for the initiation of a detonation
\citep[e.g.,][]{holcomb13,shen14}. Therefore, we extend our previous work here 
with reactive hydrodynamics calculations for a variety of merger remnants.
We map nine 3D WD-WD merger remnants representative for the entire
WD merger parameter space study of \cite{dan14} onto a 
2D axisymmetric cylindrical geometry and evolve them with the
grid-based code FLASH \citep{fryxell00}.

In this study, we expand the previous studies in several ways.
Previous multidimensional calculations have started from the initial
conditions constructed from spherical symmetric hydrostatic
equilibrium models  \citep[e.g.,][]{fink07,sim12}
or from the results of spherically symmetric accretion models  
\citep[e.g.,][]{moll13}, while our calculations are starting from
realistic WD-WD binary mergers initial conditions taken from the 
3D SPH calculations of \cite{dan14}. 
Moreover, all previous studies have neglected the angular
momentum while in our in models the merger product rotationally
supported. If the core detonation is triggered, the rotation together
with the presence of the disk can influence the morphology of the
ejecta. We carry three sets of tests for each of the nine systems:
\begin{itemize}
\item In the first set of tests, we investigate whether the WD remnants
  could lead to detonations shortly after the restart of the
  simulations with FLASH.
\item For the second set of tests, we manually setup 
  hotspots according with several criteria guided by the 3D SPH
  simulations. The reason behind this set of tests is that after the
  mapping onto the grid, the hotspots in the envelope are
  smoothed out, affecting the thermal evolution. Chemical compositions
  are also influenced, especially inside chemically mixed regions. For
  the successful detonations, we also test the impact of different
  temperature and chemical profiles inside hotspots.
\item In a third set of calculations, we asume ad hoc initial conditions in
  the hotspots following the critical conditions for a spontaneous initiation of a
  detonation from the spatially resolved 1D calculations of
  \cite{holcomb13,shen14}  for He  composition and
  \cite{roepke07,seitenzahl09} for CO.  
\end{itemize}

This paper is organized as follows. In Section \ref{sec:methods} we
briefly describe our numerical methods and the initial conditions. In Section
\ref{sec:results} we present our results in detail, with a focus on
the final nucleosynthetic yields and remnant structure for
the successfully detonating systems. 
In Section \ref{sec:summary} we summarise our results.

\section{Numerical methods}
\label{sec:methods}

To evolve the WD merger remnants, we are using the Adaptive Mesh
Refinement (AMR) FLASH application framework (version 4.2)
(\verb+http://flash.uchicago.edu+) developed 
by the University of Chicago \citep{fryxell00}. 

We use the directionally unsplit hydrodynamics solver for the
Euler equations for the compressible gas dynamics, the 
PARAMESH library \citep{macneice00} to
manage the block-structured, oct-tree adaptive grid, a multipole
solver \citep{couch13} to calculate the self-gravity of the flow and the
  Helmholtz equation of state \citep{timmes00} to 
approximate the thermodynamic properties of the matter.
To calculate the abundance changes and the nuclear energy release, a 13
  isotope $\alpha-$chain plus heavy ion nuclear 
  reaction network \citep[``aprox13'';][]{timmes99} is used containing
  $\!\,^4{\rm He}$, $\!\,^{12}{\rm 
    C}$, $\!\,^{16}{\rm O}$, $\!\,^{20}{\rm Ne}$, $\!\,^{24}{\rm
    Mg}$, $\!\,^{28}{\rm Si}$,  $\!\,^{36}{\rm S}$,
  $\!\,^{40}{\rm Ar}$, $\!\,^{40}{\rm Ca}$, $\!\,^{44}{\rm Ti}$,
  $\!\,^{48}{\rm Cr}$,  $\!\,^{52}{\rm Fe}$, $\!\,^{56}{\rm
    Ni}$ \citep{timmes99}. 
  As suggested by \cite{fryxell89,papatheodore14}, nuclear burning was 
  suppressed inside the numerically broadened shocks in order to
  ensure correct detonation speeds and associated quantities.
  In order to ensure the coupling between the hydrodynamics and
  burning, we are using a burning timestep limiter
  \citep{hawley12,kashyap15}. The burning timestep is set
  such that within a given timestep the burning energy contribution to the
  internal energy of a grid cell does not exceed a fraction of $f<1$. For
  all our calculations, we have used $f=0.5$, but we have also run
  tests with more restrictive values of 0.3 and 0.1. By reducing $f$,
  and therefore the time step, we find that the dynamics of the
  explosion does not change (e.g., second detonation occurs exactly
  at the same moment and location). The amount of
  $\!\,^{56}{\rm Ni}$ is reduced by 13\% when using $f=0.3$. This is
  larger than the difference of 3\% found by 
  \cite{kashyap15}. Using a value of 0.1, the simulations
  becomes prohibitive with the timestep reduced by four orders of
  magnitude \citep[a similar effect was seen by ][]{hawley12}. 
Our calculations do not include the burning limiter suggested by 
\cite{kushnir13}. This limiter guarantees that the burning time is 
longer than the sound crossing time of the grid cell and thus, in effect, an 
artificial numerical ignition does not occur. Comparing to
the work of \cite{hawley12}, who have not used the burning limiter,
\cite{kushnir13} have shown that the detonation ignition occurs
prematurely in the context of WD-WD collisions. This effect will be
tested in future work.

\begin{figure*}
\centerline{
 \includegraphics[height=2.15in]{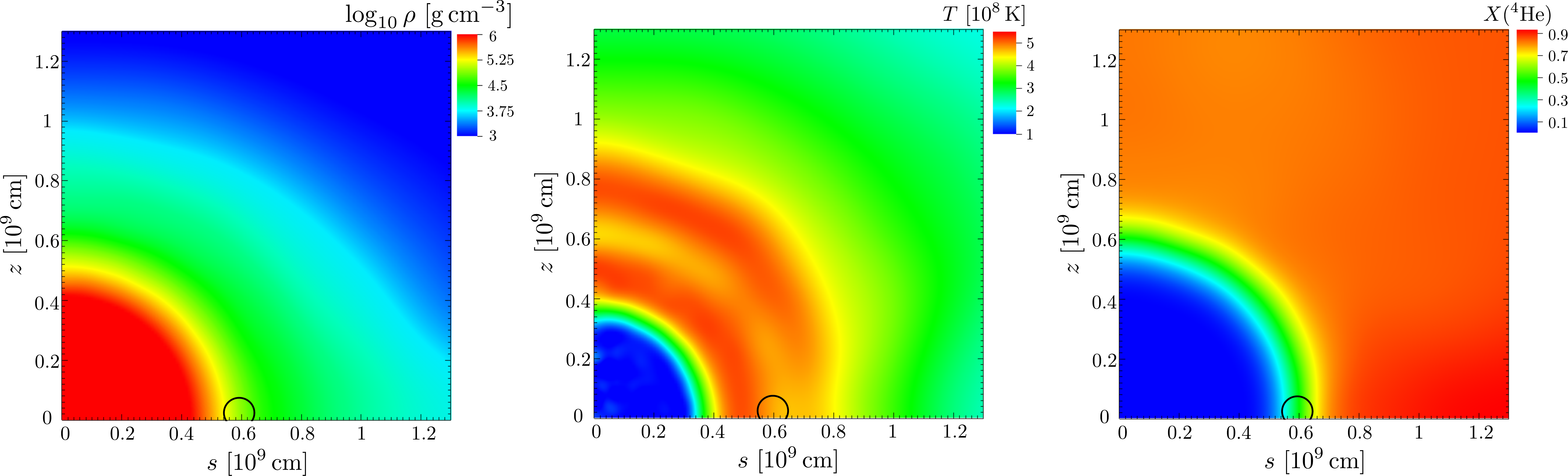}
}
\caption{Density (left), temperature (center) and the helium mass fraction (right) after the mapping 
onto the 2D cylindrical grid for the  merger remnant of an initial system with $0.45+1.1\, M_{\odot}$ 
components. For this particular system, the three search criteria ($C_1-C_3$) return a single 
hotspot, located in the first quadrant and represented with a circle of 500 km radius.} 
\label{fig:HSwd045wd110}
\end{figure*}

All calculations have been carried out in 2D cylindrical geometry,
taking rotation into account (so-called ``2.5D'' approach). 
The computational domain is divided into blocks (16 in both $s$- and
$z$-direction, where $s$ is the cylindrical radius and $z$ is the
height), each containing $[16\times 16]$ computational zones. Five
levels of refinement were allowed, yielding an effective grid size of 
$[4096\times 4096]$ or 2.7 km for a domain extending up to 
$1.1\times 10^9$ cm.
Lower and larger grid sizes are also used, between $1.0 - 1.5\times 10^9\,
{\rm cm}$, depending on the mass ratio between the donor and the
accretor, $q=M_{2}/M_{1}$: lower/higher would produce a more/less
extended remnant.

For the models where a detonation is ignited, we rerun the simulations
with an extended computational domain of $s_{\rm max}= z_{\rm max}=
10^{11}$ cm. This allows us to follow the nucleosynthesis processes
until the ejecta reaches the homologous phase. For these runs, we have
changed the grid parameters to nine levels of refinement and
$[32\times 32]$ computational zones/block in order to yield an
effective grid resolution of 7.6 km.

\subsection{Initial models}
\label{sec:ICs}

For our initial models, we use the WD-WD parameter space studies of 
\cite{dan12,dan14}.
We follow the evolution of nine systems, representing all possible
chemical composition combinations in the WD-WD parameter
space. 
We choose the following WD compositions: pure He for $M_{\rm WD}\leq
0.45\, M_\odot$; a hybrid composition consisting of an CO ($X[\!\,^{12}{\rm C}]=0.5$,
$X[\!\,^{16}{\rm O}]=0.5$) core surrounded by a pure He envelope of
$0.1\, M_\odot$ for $0.45\,M_\odot < M_{\rm WD} \leq 0.6\, M_\odot$;
pure CO ($X[\!\,^{12}{\rm C}]=0.4$, $X[\!\,^{16}{\rm O}]=0.6$) for
$0.6\,M_\odot < M_{\rm WD} \leq 1.05\, M_\odot$ and made of ONe and
magnesium (Mg)
($X[\!\,^{16}{\rm O}]=0.6,\ X[\!\,^{20}{\rm Ne}]=0.35$ and
$X[\!\,^{24}{\rm Mg}]=0.05$) for $M_{\rm WD}\geq 1.05\, M_\odot$.  
As the maximum temperature inside the remnant tends to increase with the total mass 
\citep{zhu13,dan14}, from each of these nine regions, we choose one of
the most massive WD-WD remnants:
\begin{itemize}
\item[$\mathbf{1)}$]  $\mathbf{0.45 + 0.45\,{M}_\odot}$ -- largest total mass of two WDs that are
  made of He;
\item[$\mathbf{2)}$]  $\mathbf{0.45 + 0.6\,M_\odot}$ -- largest mass combination of a WD made
  of He and a hybrid one, made of He and CO;
\item[$\mathbf{3)}$]  $\mathbf{0.45 + 0.9\,M_\odot}$ -- largest He with a massive CO;
\item[$\mathbf{4)}$]  $\mathbf{0.45 + 1.1\,M_\odot}$ -- largest He with a massive WD made of
  ONe;
\item[$\mathbf{5)}$]  $\mathbf{0.55 + 1\,M_\odot}$ -- massive hybrid He-CO WD with massive CO WD;
\item[$\mathbf{6)}$]  $\mathbf{0.6 + 0.6\,M_\odot}$ -- most massive hybrid He+CO WD system;
\item[$\mathbf{7)}$]  $\mathbf{0.6 + 1\,M_\odot}$ -- massive hybrid He-CO WD with a massive WD made of ONe;
\item[$\mathbf{8)}$]  $\mathbf{0.95 + 1.15\,M_\odot}$ -- massive CO plus massive WD made of ONe;
\item[$\mathbf{9)}$]  $\mathbf{1.05 + 1.05\,M_\odot}$ -- largest total mass of two CO WDs.
\end{itemize}

The 3D WD-WD merger remnants from \cite{dan14} are mapped 
onto a 2D axisymmetric cylindrical Eulerian mesh, with the averaging done
over the azimuthal angle $\phi$. 
The snapshots are taken at a time of three initial orbital
periods after the moment when the donor is fully disrupted. At this
stage the dynamical evolution is essentially over.
The result of the mapping can be seen in Figure \ref{fig:HSwd045wd110}
for the $0.45+1.1\, M_{\odot}$ system.

We run three sets of simulations:
\begin{enumerate}
\item In a first set, we investigate whether the WD remnants
could lead to detonations shortly after the restart of the simulations
with FLASH, without any change of their properties (i.e., no
perturbations). Runs are ordered by increasing donor's mass and by
decreasing mass ratio q and labeled 1 through 9.
The main properties of this set of runs are listed in Table
\ref{tab:ICsnoper}.  
\item Through mapping, there is a 
loss of resolution because first, the SPH
interpolation is required to get the values on the grid 
and second, an averaging is done over the azimuthal angle. This has an impact
onto the thermal evolution of the remnant as the hotspots
are smoothed out and chemical compositions change, especially inside
chemically mixed regions. 
This motivates the second set of runs  (initial conditions are
presented in Table \ref{tab:ICs}), where we manually setup hotspots
based on different criteria guided by the 3D SPH 
  simulations: ${C_{1}}$ uses the ratio
  between the nuclear and dynamical timescales $\tau_{\rm
    nuc}/\tau_{\rm dyn}(T)$ \citep[effective timescales are computed
  as in][]{dan12}; 
${C_{2}}$ uses the maximum temperature over all
  particles $T_{\rm max}$, and ${C_3}$ uses the maximum temperature over particles
  above a threshold density of $10^5\, {\rm g\, cm^{-3}}$,
  $T_{\rm max}(\rho\geq {10^5}\, {\rm g\,cm^{-3}})$, respectively.
For CO mass-transferring systems besides criteria $C_1$ and $C_2$, we
also use $T_{\rm max}(\rho\geq  10^6\, {\rm g\, cm^{-3}})$  (${C_{4}}$).
For this set of runs, all SPH particles are searched to locate the one
satisfying the condition specified by the selection criteria. Its
position $(s_{\rm per},z_{\rm per})$ and temperature ($T_{\rm per}$) are used to
setup the hotspot, see below for more details. Models are ordered in
the same way as for the previous set, only that after the run number
the search criteria ($C_1-C_4$) was added. For example, run 4
($0.45+1.1\, M_{\odot}$) using the results of the hotspot search
criteria $C_1$ was named ``4.c1''.
\item Finally, for the third set of tests (see Table \ref{tab:ICsdd}),
the hotspots are setup following the critical conditions for a
spontaneous initiation of a detonation from the spatially resolved
1D calculations of \cite{holcomb13} and \cite{shen14} for He composition and
\cite{roepke07} and \cite{seitenzahl09} for CO. For this set of runs,
we assume that the systems have evolved to reach the conditions
necessary for spontaneous detonations determined in the previously
cited studies, as the spatial scales relevant for the initiation of
  the He or CO detonations cannot be resolved in our multidimensional
  simulations \citep[He and C detonation length scales, at densities
  above $\rho=10^7\, {\rm g\, cm^{-3}}$, are below $10^4$ cm and 1 cm,
  respectively;][]{holcomb13,shen14a}. 
For He mass transferring tests we setup hotspots with the following
properties: a temperature 
perturbation of $5\times 10^8$ K (or $8\times 10^8$ K, if the
hot envelope temperature is close to $5\times 10^8$ K) at densities above
$3-5\times 10^5\, {\rm g\,cm^{-3}}$. This is
what sets a direct detonation below a perturbation radius of about
$R_{\rm per}=1000$ km \citep{holcomb13,shen14}. 
For the CO mass-transferring systems, we use $T_{\rm per}=3\times
10^9$ K at densities above $\rho=1\times 10^6\, {\rm g\,cm^{-3}}$.
Models are ordered in the same way as above, with the label ``dd'' (an
abbreviation to indicate ``direct detonation'') added after the run number.
\end{enumerate}

\cite{seitenzahl09} explored the effect of different initial
profiles for density, temperature and chemical composition 
on the spontaneous initiation of CO detonations and 
found that the different profiles change the hotspots critical sizes.
For the runs with perturbations, we also test the impact of the
hotspot temperature and composition profiles  (see \S \ref
{sec:profiles} and \ref{sec:HSsize}), and, because our
calculations are in 2D cylindrical geometry, the impact of hotspot
geometry (see \S \ref{ref:hsgeom}).
While we carefully choose where to place the hotspots and test
the different setups, further explorations are required as several
assumptions are made here. For example, it is not clear how
the ignition of a detonation may depend on the time dependence of
the hotspot formation. Our setups for the second and
third set of tests is based on the 
main assumption that hotspots with properties (presented in Section
\ref{sec:results}) close to those leading to the core  detonation
develop during the WD merger remnants evolution. 
The second set of tests is based on a realistic approach, but we
  still have
  modified the hotspots temperature profiles, both, in extent and shape.  
  For the third set, the assumptions were further relaxed, with
  hotter hotspots placed in denser regions of the
  envelope. Probably, such conditions could be realised only 
  if the initial spin state of the WDs would be different. For
  example, it has been shown in \cite{zhu13,dan14} that starting with
  non-rotating initial stars the location of peak temperature is
  located in a denser environment compared to when starting with
  corotating stars, like in the present study. The effect of the
  initial conditions will be investigated in a future study.

\subsection{Hotspot size}
\label{sec:HSsize}

For the runs with perturbations (second and third sets)
our strategy is to start with a relatively large perturbation radius
inside the shell of $R_{\rm per} = 1000$ km. Such initial large
  hotspots are unrealistic and are not seen in the SPH simulations of
  WD mergers and they are more characteristic to WD collisions 
\citep{rosswog09c,raskin09,aznar13,kushnir13,garcia13,papish15}. The
  SPH interaction radius of the particles inside the hot envelope is below
  $\sim\! 1000$ km, decreasing with increasing density, towards the
  core where the hotspots are located. If a detonation is
triggered into the underlying core, we repeat the 
simulations with a new hotspot radius $R_{\rm per}$ which is
determined by the bisection root-finding method.
We repeat this process until the core does not
detonate anymore and the difference between the hotspot radius
of a successful and failed run is below $15\%$. The minimum 
perturbation radius $R_{\rm per}$ at which the core detonates is
included in Tables \ref{tab:results2} and \ref{tab:results3}.
We note that the minimum value thus obtained is method- and 
ICs-dependent and can change with the star's spin and chemical
composition, numerical resolution and the nuclear reaction network
(see a detailed discussion of these factors in Section \ref{sec:results}).

\subsection{Hotspot temperature profile}
\label{sec:profiles}

The initial temperature profile from the SPH simulations can take
different shapes. If not specified otherwise, the hotspots are setup
with a flat temperature profile, i.e., constant temperature $T_{\rm
  per}$ inside the perturbation radius $R_{\rm  per}$. 
To test the dependence on the temperature profile, in addition to the
flat temperature profile, we setup hotspots using a linear profile
\begin{equation}\label{eq:lprof}
T(r)= 
\left\{
\begin{array}{ll}
T_{\rm per} - \frac{T_{\rm per}-T_{\rm amb}}{R_{\rm per}} r & \textrm{for $ r \le R_{\rm per}$,}\\
\\
T_{\rm per} &\textrm{for $r > R_{\rm per}$,}
\end{array} \right.
\end{equation}
as well as Gaussian
\begin{equation}\label{eq:gprof}
T(r) = T_{\rm amb} + (T_{\rm per} - T_{\rm amb})  e^{-(r/R_{\rm per})^2},
\end{equation}
where $r$ is the distance from the hotspot origin and $T_{\rm amb}$
is the ambient temperature, chosen as the highest
temperature in the neighbour cells, surrounding the hotspot.

\subsection{Hotspot composition}
\label{sec:hscomp}

The mixing of matter between the binary components
causes the composition to be inhomogeneous
mainly through the outer layers of the core, the hot envelope and the
inner disk. It occurs mainly during the last phase of the merging
process and it tends to increase with the
mass ratio $q=M_{\rm   don}/M_{\rm   acc}$ \citep{zhu13,dan14}. 
Only two regions of the parameter study have both stars made of
the same fuel: He--He and CO--CO. The other regions are made of a
combination of He, CO and ONe. 

The third panel of Figure \ref{fig:HSwd045wd110} shows the
mass fraction of He for a $0.45+1.1\,
M_\odot$ system. For this system, the hotspot search criteria $C_1$
returns a particle with mainly CO neighbour particles (i.e., particles
contributing to the interpolation), but also $\sim\! 15\%$ made of
pure He. 
Similar to the hotspot temperature, discussed in the previous
section, after the mapping onto the 2D grid some details about the
mixing are lost and, in some cases, the fuel (He or CO) concentration
is reduced to very low numbers. Specifically, for the models with 
$0.55+1\, M_\odot$ and $0.6+1.1\, M_\odot$ initial components, there is a
lack of He fuel ($X[\,\!^{4}{\rm He}]<0.1$) inside the hot envelope
at densities above $\rho\geq 6\times 10^4\, {\rm g\,  cm^{-3}}$ 
and $\rho\geq 3.9\times 10^4\, {\rm g\, cm^{-3}}$, respectively.
For these two systems we have set the hotspot composition to pure He.

\subsection{Hotspot geometry}
\label{ref:hsgeom}

In 2D cylindrical geometry, the hotspots are tori extending around
the star. For the $0.45+0.6\,M_\odot$ and
$0.45+0.9\,M_\odot$ systems we have also setup spherical hotspots,  
i.e., the perturbation was placed at the pole, i.e., where $s=0$ (run
2.dd-Polar and 3.dd-Polar, respectively, see Table \ref{tab:results3}).  
Also for these two tests, we determine the minimum perturbation radius
leading to a (core) detonation through bisection.

\section{Results}
\label{sec:results}

\subsection{Models without perturbation}

\begin{center}
\begin{small}
\begin{table*}
\begin{tabular}{|c|c|c|c|c|c|c|c|c|}
  \hline\hline
Run & Initial & \multirow{2}{*}{$T_{\rm max,8}$} & \multirow{2}{*}{$\rho_5(T_{\rm max})$} 
    & \multicolumn{2}{|c|}{Ignition} & \multirow{2}{*}{$t_{\rm det.env,s}$} & \multirow{2}{*}{$t_{\rm det.core,s}$} &
\multirow{2}{*}{Comments}\\*
  \cline{5-6}
    No & masses [$M_\odot$] & & &  Envelope &  Core & & & \\
    \hline\hline
\multicolumn{9}{|c|}{Helium mass transferring systems}\\
    \hline
  1  & $0.45+0.45$ & 2.786 & 0.009 & No & No & - & - & $\rho$(hot envelope) too low, no burning\\
  \hline
  2  & $0.45+0.6$ & 2.822 & 0.051 & No & No & - & - &  $\rho$(hot envelope) too
                                           low, no burning \\
  \hline
  3 & $0.45+0.9$ & 5.781 & 0.358 & No & No & - & - &  $\rho$(hot envelope) too
                                           low, no burning \\
  \hline
\multirow{2}{*}{4}  & \multirow{2}{*}{$0.45+1.1$} & \multirow{2}{*}{19.581} &
                                                   \multirow{2}{*}{2.007}
    & \multirow{2}{*}{Yes} & \multirow{2}{*}{No} & \multirow{2}{*}{1.3} &
                                                                       \multirow{2}{*}{-}
       & core intact, envelope burns mainly\\
        & & & & & & & & to IMEs and some IGEs\\
  \hline
\multirow{2}{*}{5} & \multirow{2}{*}{$0.55+1.0$} & \multirow{2}{*}{6.556} &
                                                                      \multirow{2}{*}{1.041} & \multirow{2}{*}{No} & \multirow{2}{*}{No} & \multirow{2}{*}{-} & \multirow{2}{*}{-} & no burning, very
                                             little He left inside \\
& & & &  & & & & hot envelope after mapping \\
  \hline
6 & $0.6+0.6$ & 4.035  & 1.120 & No & No & - & - & little burning inside
                                              hot envelope\\
\hline
\multirow{3}{*}{7} & \multirow{3}{*}{$0.6+1.1$} & \multirow{3}{*}{8.158}  &
                                                                       \multirow{3}{*}{1.165}&
                                                                                               \multirow{3}{*}{No}
                                     & \multirow{3}{*}{No} &
                                                             \multirow{3}{*}{-}
                                                     &
                                                       \multirow{3}{*}{-}
       & temperature inside hot envelope high, 
  \\
 &  & & & & & & & but little He there ($X(\,\!^{4}{\rm He})_{\rm max}\!\approx\!
                        10^{-3}$); \\
&  & & & & & & & some burning (small
                 amount of IMEs)\\
  \hline 
  \multicolumn{9}{|c|}{Carbon/oxygen mass transferring systems}\\
  \hline
        8 & $0.95+1.15$ & 11.561 & 9.493 & No & No & - & - & temperature too low; little burning\\ 
\hline
9 & $1.05+1.05$ & 11.172 & 0.731 & No & No & - & - & temperature too low; little burning\\
\hline\hline
\end{tabular}
\caption{{Outcomes for the first set of runs, without perturbation}:
  masses ($M_\odot$) at the beginning of mass transfer, when the
  simulations are initialised in SPH; maximum
    temperature $T_{\rm max,8}$ (in units of $10^8$ K) and density at
    the location of maximum temperature $\rho_5(T_{\rm max,9})$ (in
    units of $10^5\, {\rm g\, cm^{-3}}$) during the entire evolution;
    $t_{\rm det.env,s}$ and $t_{\rm det.core,s}$ are the time (in
    units of s) after which a detonation is ignited in the envelope
    and in the core, respectively.} 
\label{tab:results1}
\end{table*}
\end{small}
\end{center}

A summary of the outcome for this set of runs is presented in Table \ref{tab:results1}.
Only the $0.45+1.1\,M_\odot$ system triggers a (He) shell detonation. Shocks propagate 
through the underlying (ONe) core, but the temperature is not hot
enough to fuse oxygen. 
The nucleosynthetic yields for this run are shown in the fourth column
of Table \ref{tab:nuc}, in the Appendix. The  
  yields are low and the majority of burning  products are in the form
  of intermediate-mass elements (IMEs; with the atomic mass number $A$
  ranging from 28 to 40), with very low yields of iron-group elements
  (IGEs; with $A\geq 44$). Most of the He remains unburned 
  (initial He mass of $M(\,\!^{4}{\rm He})=0.45\, M_\odot$). This will
  be a very faint event, much fainter than a ``point'' Ia supernova
  \citep{shen10}. 
The main reasons why the other systems containing He do not trigger a
detonation without a perturbation is that the hot regions (i.e.,
envelope) in some of our merger remnants have typical densities of
$\rho \sim 10^5\, {\rm g\,  cm^{-3}}$ and detonations are not 
expected to be ignited at this density. \cite{woosley11},  assuming that 
the accreted material is made of 99\% $\,\!^{4}{\rm He}$ and 1\%
$\,\!^{14}{\rm N}$, have shown that He does not directly ignite a
detonation below a density of $\sim 10^6\, {\rm
  g\,cm^{-3}}$, unless the hotspot size is a significant fraction of
the WD scale height.  \cite{holcomb13} reached a similar conclusion
assuming a pure He composition of the burning region. 
However, the envelopes in our models, containing initially a (hybrid) He
mass-tranferring star and a (hybrid) CO accretor, are not made of pure
He, but a mixture of He and CO. It has been shown that by adding
a fraction of CO to the He hotspot, depending on the temperature,
the hotspot size could be reduced by  more than an order of magnitude
\citep{shen14}. The hotspot size could be further reduced if instead
of using the ``aprox13'' network, we would use a large reaction
network in order to allow for the $(p,\gamma)(\alpha,p)$, with the
proton playing a catalytic role near the beginning of the
$\alpha$-chain \citep{woosley11,shen14}. We defer a study of the
effects of a larger network to future research. 

For the CO systems,  we have found that after the mapping onto the 2D
grid, the systems have temperatures below the threshold for
carbon-ignition. For those with a hybrid He+CO former donor
($0.55+1\, M_\odot$ and $0.6+1.1\, M_\odot$) there is a lack of He
fuel inside the hot envelope. While temperatures and densities are
high inside the hot envelope (see Table \ref{tab:ICsnoper}), the He
mass fraction is very low $X(\,\!^{4}{\rm He})<0.1$.

\begin{center}
\begin{small}
\begin{table*}
\setlength{\tabcolsep}{5.5pt}
\begin{tabular}{|c|c|c|c|c|c|c|c|c|c|}
  \hline\hline
Run & Initial & \multirow{2}{*}{$R_{\rm per,km}$} & \multirow{2}{*}{$T_{\rm max,8}$} & \multirow{2}{*}{$\rho_5(T_{\rm max})$} 
    & \multicolumn{2}{|c|}{Ignition} & \multirow{2}{*}{$t_{\rm det,s}$} & \multirow{2}{*}{$t_{\rm det.core,s}$} &
\multirow{2}{*}{Comments}\\*
  \cline{6-7}
    No & masses [$M_\odot$] & & & &  Envelope &  Core & & & \\
    \hline\hline
\multicolumn{10}{|c|}{Helium mass transferring systems}\\
    \hline
  1.c1 & \multirow{3}{*}{$0.45+0.45$} & 1000 & 2.786 & 0.009 & No & No
                                                                        & - & - & \multirow{1}{*}{$\rho$(hot envelope) too low}\\
  \cline{1-1} \cline{3-9}
  1.c2 & & 1000 & 2.920 & 0.009 & No & No & - & - & no burning \\
  \cline{1-1} \cline{3-9}
  1.c3 & & 1000 & 2.781 & 0.009 & - & - & - & - & \\
  \hline
  2.c1 & \multirow{3}{*}{$0.45+0.6$} & 1000  & 2.760 & 0.061
    & No & No & - & - &  \multirow{1}{*}{ little He burning in }\\
  \cline{1-1} \cline{3-9}
  \multirow{1}{*}{2.c2} & & \multirow{1}{*}{1000} & \multirow{1}{*}{3.353} &
                                                   \multirow{1}{*}{0.130}
    & \multirow{1}{*}{No} & \multirow{1}{*}{No} & \multirow{1}{*}{-} &
                                                                       \multirow{1}{*}{-}
       &  and around hotspot \\
  \cline{1-1} \cline{3-9}
  \multirow{1}{*}{2.c3} & & \multirow{1}{*}{1000} & \multirow{1}{*}{2.838} &
                                                   \multirow{1}{*}{0.042}
    & \multirow{1}{*}{No} & \multirow{1}{*}{No} & \multirow{1}{*}{-} &
                                                                       \multirow{1}{*}{-}
       &  \\
  \hline
 3.c1 & \multirow{3}{*}{$0.45+0.9$} & \multirow{2}{*}{1000} & \multirow{2}{*}{6.857} &
                                                   \multirow{2}{*}{0.462}
    & \multirow{2}{*}{No} & \multirow{2}{*}{No} & \multirow{2}{*}{-} &
                                                                       \multirow{2}{*}{-}
       &  envelope perturbed; \\
3.c3 & & & & & & & & & He burning in and around\\
  \cline{1-1} \cline{3-9}
  \multirow{1}{*}{3.c2} & & \multirow{1}{*}{1000} & \multirow{1}{*}{6.294} &
                                                   \multirow{1}{*}{0.448}
    & \multirow{1}{*}{No} & \multirow{1}{*}{No} & \multirow{1}{*}{-} &
                                                                       \multirow{1}{*}{-}
       &  hotspot region \\
  \hline
  4.c1 & \multirow{4}{*}{$0.45+1.1$} & \multirow{3}{*}{500}  &\multirow{3}{*}{67.758} &
                                                          \multirow{3}{*}{801.038}
    & \multirow{3}{*}{Yes} & \multirow{3}{*}{Yes} & \multirow{3}{*}{0.34} &
                                                                            \multirow{3}{*}{1.74}
       & core ignited off-center and burns \\
  4.c2   & & & & & & & & & up to $\,\!^{56}{\rm Ni}$; envelope  burns 
                           \\
    & & & & & & & & & mainly to IMEs, little to IGEs\\
  \cline{1-1} \cline{3-10}
  4.c3 & & -  & - & - & - & -& -& - & no run as $T_{\rm per}\approx T_{\rm grid}$ \\
  \hline
  5.c1 & \multirow{3}{*}{$0.55+1.0$} & \multirow{3}{*}{1000} & \multirow{3}{*}{10.393} &
                                                                      \multirow{3}{*}{1.491}
    & \multirow{3}{*}{No} & \multirow{3}{*}{No} & \multirow{3}{*}{-} &
                                                                       \multirow{3}{*}{-}
       & little burning (no IGEs) around 
         \\
5.c2 & & & & & & & & & hotspot region; initial \\
5.c3 & & & & & & & & &$X(\,\!^{4}{\rm He})_{\rm max}\!\approx\!
                        10^{-3}$  inside hotspot\\
  \hline
6.c1 & \multirow{3}{*}{$0.6+0.6$} & \multirow{3}{*}{1000} & \multirow{3}{*}{3.951} &
                                                                      \multirow{3}{*}{0.880}
    & \multirow{3}{*}{No} & \multirow{3}{*}{No} & \multirow{3}{*}{-} &
                                                                       \multirow{3}{*}{-}
       & envelope perturbed by initial \\
  6.c2 & & & & & & & & & wave-like shocks, but only \\
  6.c3 & & & & & & & & & little burning going on \\
  \hline
7.c1 & \multirow{4}{*}{$0.6+1.1$} & \multirow{2}{*}{1000} & \multirow{2}{*}{8.085}  &
                                                                       \multirow{2}{*}{1.962}&
                                                                                               \multirow{2}{*}{No}
                                     & \multirow{2}{*}{No} &
                                                             \multirow{2}{*}{-}
                                                     &
                                                       \multirow{2}{*}{-}
       & \multirow{2}{*}{similar evolution with run 7}\\
7.c3 & & & & & & & & & \\
\cline{1-1} \cline{3-10}
\multirow{2}{*}{7.c2} & & \multirow{2}{*}{1000} & \multirow{2}{*}{8.125}  &
                                                                       \multirow{2}{*}{1.179}&
                                                                                               \multirow{2}{*}{No}
                                     & \multirow{2}{*}{No} &
                                                             \multirow{2}{*}{-}
                                                     &
                                                       \multirow{2}{*}{-} &  evolution similar
         with run 7 \\
& & & & & & & & & but envelope more
         perturbed\\
\hline
  \hline 
  \multicolumn{10}{|c|}{Carbon/oxygen mass transferring systems}\\
  \hline
  \multirow{2}{*}{8.c1} & \multirow{5}{*}{$0.95+1.15$} & \multirow{2}{*}{1000} & \multirow{2}{*}{12.798}
                                                                               & \multirow{2}{*}{6.607} & \multirow{2}{*}{No} & \multirow{2}{*}{No} & \multirow{2}{*}{-} & \multirow{2}{*}{-} & envelope perturbed; little burning \\
  & & & & & & & & & inside envelope \\
\cline{1-1} \cline{3-10}
8.c2 & & \multirow{2}{*}{1000} & \multirow{2}{*}{13.004}
                                                                               &
                                                                                 \multirow{2}{*}{10.876}
    & \multirow{2}{*}{No} & \multirow{2}{*}{No} & \multirow{2}{*}{-} &
                                                                       \multirow{2}{*}{-}
       & \multirow{2}{*}{similar evolution with run 8.c1}\\
8.c4  & & & & & & & & & \\
  \hline 
9.c1 & \multirow{3}{*}{$1.05+1.05$} & \multirow{3}{*}{1000} & \multirow{3}{*}{10.414} &
                                                                       \multirow{3}{*}{64.259}
    & \multirow{3}{*}{No} & \multirow{3}{*}{No} & \multirow{3}{*}{-} &
                                                                       \multirow{3}{*}{-}
       & similar with run 9; \\
9.c2 & & & & & & & & & burning  only inside hotspot's \\
9.c4 & & & & & & & & & higher  density  region\\
\hline\hline
\end{tabular}
\caption{Outcomes for the second set of runs, with perturbation based
  on different criteria guided by the 3D SPH simulations: same
  quantities as in Table \ref{tab:results1}, only the hotspot radius
  $R_{\rm per,km}$ (in units of km) has been added.} 
\label{tab:results2}
\end{table*}
\end{small}
\end{center}

\subsection{Models with perturbation based on SPH calculations guided criteria}
\label{sec:rest2}

\begin{figure*}
\centerline{
 \includegraphics[height=9.2in]{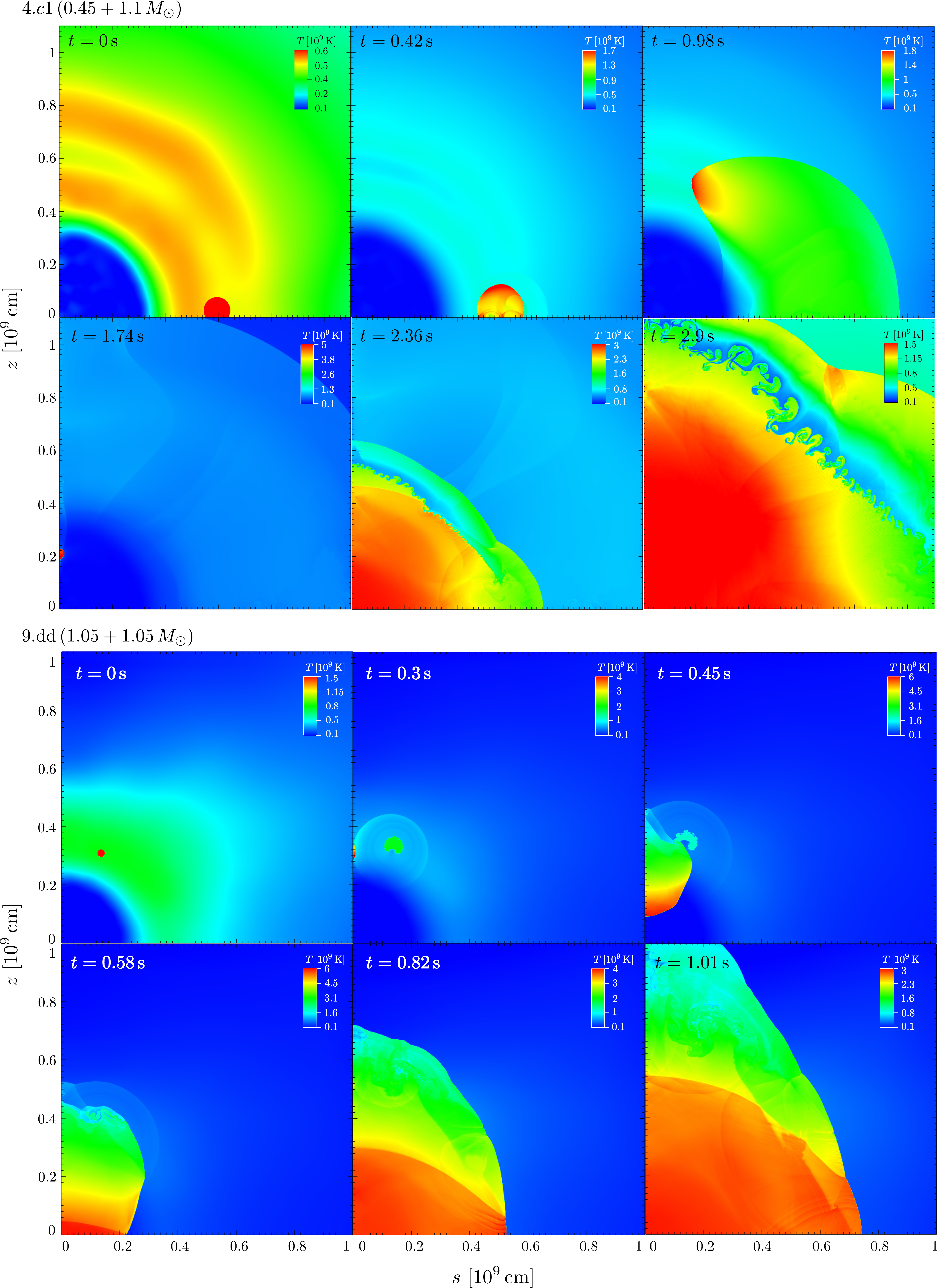}
}
\caption[Evolution of the systems with 
  $0.45+1.1\, M_{\odot}$ (run number 4.c1; upper panels) and
  $1.05+1.05\, M_{\odot}$ (run number 9.dd; lower panels)
  components.]{Evolution of the 
  $0.45+1.1\, M_{\odot}$ (run number 4.c1; upper panels) and
  $1.05+1.05\, M_{\odot}$ (run number 9.dd; lower panels) systems. 
  Color coded is the temperature in units of $10^9\, {\rm K}$. 
}
\label{fig:tempevol}
\end{figure*}

A summary of the outcome for this set of runs is presented in Table
\ref{tab:results2}. 
With perturbations set based on the hotspot search criteria guided by
the 3D SPH simulations presented in Section \ref{sec:ICs}, again, only
the $0.45+1.1\,M_\odot$ system detonates (run 4.c1).  For this system,  
the criteria $C_1$ and $C_2$ return the same hotspot, close to the orbital
plane ($s=0.595\times 10^9$ cm and $z=0.026\times 10^9$ cm). The
difference, compared to the model without perturbation, run $4$, is that 
in this case also the (ONe) core detonates, see the upper panels of
Figure \ref{fig:tempevol}.
As discussed in Section \ref{sec:HSsize}, we start with a relatively
high perturbation radius and then decrease it through successive
bisections, until the difference between hotspot radius of a succesful
and a failed run is below $15\%$. 
For this case, the minimum perturbation radius at which the second
detonation is triggered is $R_{\rm per}= 500$ km.  
Above this value, the evolution is the same: triple alpha reactions
quickly raise the temperature to $\sim 1.9\times 
10^9$ K so that a large overpressure is produced. A self-sustained
detonation is formed, which wraps around the (ONe) core and
converges inside the core close to the polar axis (due to the cylindrical
symmetry of our models). At that point, the shock wave is sufficiently strong
to cause a shock-initiated off-center core detonation.
The ONe detonation occurs when a spherical region ($z=1950$ km and
$r\approx 0$) of a $(100\ {\rm km})$ radius reaches a temperature of
$T\approx 6\times 10^9$ K at a density of $\rho\approx 4\times 10^7\,
{\rm g\, cm^{-3}}$. The robustness of the ONe core detonation is
further discussed in Section \ref{sec:nuc}. 

For all He mass-transferring systems, we have run two extra sets of
simulations with an increased density threshold of $\rho \geq 5\times
10^5\, {\rm g\, cm^{-3}}$ and $\rho \geq 1\times 10^6\, {\rm g\,
  cm^{-3}}$, but without success. Above these thresholds the
temperatures are either too low or there is no or very little He
inside the hotspots.  We also run tests using multiple hotspots for
the systems returning 
different hotspots criteria: $0.45+0.9\,M_\odot$ with two hotspots set
according to the results of the search criteria $C_1$ ($C_2$ returns
the same hotspot) and $C_3$ and $0.95+1.15\,M_\odot$ with two
hotspots based on $C_1$ and $C_2$ ($C_4$ returns the same
hotspot). Also for these runs a detonation was not ignited.

\begin{center}
\begin{small}
\begin{table*}
  \begin{tabular}{|c|c|c|c|c|c|c|c|c|}
    \hline\hline
   Run & Initial  & \multirow{2}{*}{$R_{\rm per,km}$} & \multirow{2}{*}{$T_{\rm max,9}$} & \multirow{2}{*}{$\rho_6(T_{\rm max})$} 
   & \multicolumn{2}{|c|}{Ignition} & \multirow{2}{*}{$t_{\rm det.env,s}$}
   & \multirow{2}{*}{$t_{\rm det.core,s}$} \\
 \cline{6-7}
    No & masses [$M_\odot$] & & &  & 
       Envelope &  Core & & \\
    \hline\hline
    \multicolumn{9}{|c|}{Helium mass transferring systems}\\
    \hline
  1.dd & $0.45+0.45$ & 437.5 & 4.297 & 4.656 & No & Yes & - & 2.2 \\
\hline
  2.dd & $0.45+0.6$ & 875 & 8.262 & 56.471 & Yes & Yes & 0.3 & 0.3 \\
\hline                                                       
2.dd-Polar & $0.45+0.6$ & 156.25 & 8.211 & 53.964 & Yes & Yes & 0.11 & 0.2 \\
\hline      
3.dd & $0.45+0.9$ & 187.5 & 5.682 & 21.966 & Yes & Yes & 0.1 & 1.8 \\
\hline   
3.dd-Polar & $0.45+0.9$ & 750 & 5.856 & 30.337 & Yes & Yes & 0.17 & 2.7 \\
\hline   
5.dd & $0.55+1$ & 500 & 6.435 & 36.552 & Yes & Yes & 0.1 & 0.24 \\
\hline
6.dd & $0.6+0.6$ & 375 & 3.188 & 4.740 & No & Yes & - & 1 \\
\hline
7.dd & $0.6+1.1$ & 1000 & 5.178 & 21.104 & Yes & No & 0.1 & - \\
    \hline
    \multicolumn{9}{|c|}{Carbon/oxygen mass transferring systems}\\[1.0ex]
    \hline
\multirow{1}{*}{8.dd} & $0.95+1.15$ & \multirow{1}{*}{500} & \multirow{1}{*}{9.447} & \multirow{1}{*}{112.052} & \multirow{1}{*}{No} & \multirow{1}{*}{Yes} & \multirow{1}{*}{-} & \multirow{1}{*}{0.33} \\
\hline
9.dd & $1.05+1.05$ & 125 & 6.228 & 40.736 & No & Yes & - & 0.3 \\
    \hline\hline                                        
\end{tabular}
\caption{Outcomes for the third set of runs, where the hotspots are setup
following the critical conditions for a direct initiation of a
detonation from the spatially resolved 1D calculations. Same  
quantities as in Table \ref{tab:results2}, only that the temperare is in units of $10^9$ K
  and density in units of $10^6\, {\rm g\, cm^{-3}}$.}
\label{tab:results3}
\end{table*}
\end{small}
\end{center}

The fifth column of Table \ref{tab:nuc} in the Appendix shows the nucleosynthetic yields for this
  run and the upper panels of Figure \ref{fig:isotwosyst} show the
  final abundance distribution in velocity 
  space. This model should resemble  a ``normal'' SN Ia. The isotope yields are 
  very similar to those of \cite{moll13} for the system with
  $0.81+0.96\,M_\odot$ (see their Table 3) and with those of
  \cite{pakmor12} for their $0.9+1.1\,M_\odot$ \cite[see Section 3
  in][]{pakmor12}. However, our model leads to a different ejecta
  morphology compared with that  
  of \cite{pakmor12} and \cite{moll13}. While they find IMEs
  and unburned oxygen at the center of the ejecta and $\,\!^{56}{\rm
    Ni}$ above this region (see their Figure 9), our model leads to a
  SN Ia-like ejecta structure \citep{mazzali07}. 
  Similar to \cite{moll13}, the ejecta in our model is asymmetric with a
  higher velocity along the polar regions compared to the equatorial
  ones. This should lead to orientation effects for both light-curves
  and spectra, see the detailed discussion of synthetic observables in
  Section 4 of \cite{moll13}.  
  The reasons that the other systems fail to detonate are the same as
  presented above, for the runs without  
  perturbation: the hotspots temperature and/or density are either too
  low or there is no or very little He inside the hotspots.

\subsection{Models with perturbation based on 1D spatially resolved calculations}

Detonations are triggered for all runs starting with hotspots following the
critical conditions for a spontaneous initiation of a detonation from
the spatially resolved 1D calculations of \cite{holcomb13,shen14}  for
He composition and \cite{roepke07,seitenzahl09} for CO.  

Before we discuss the models in detail, we outline some of the general findings,
characteristic to the majority of the models.
In general, the second detonation happens either through a
smooth transition from the envelope into the core (edge-lit
detonation scenario) or a second detonation occurs after shock
convergence into the core (off-center detonation scenario). 
However, not all systems undergo double detonations. For run 1.dd
($0.45+0.45\,M_\odot$), the hotspot is set inside the core. This setup system
is very improbable, at least for the initial conditions (tidally
locked stars) used in the calculations of \cite{dan12,dan14}. Because the
hot envelope has a low density ($\rho_{\rm env}<10^5\, {\rm g\, cm^{-3}}$), in order to
trigger a detonation we have  
to setup a hotspot inside the cold core, in the
equatorial plane, $\approx 5400$ km above the core's center (1500 km
below the location of $T_{\rm max}$).  
The evolution is very similar for runs 6.dd ($0.6+0.6\,M_\odot$)
and 9.dd ($1.05+1.05\,M_\odot$). The perturbation does not lead to
a shell detonation, but to a sub-sonic shock wave which reflects at the
pole, at the core's edge and triggers a direct detonation there.

\begin{figure*}
\centerline{
\includegraphics[height=8.5in]{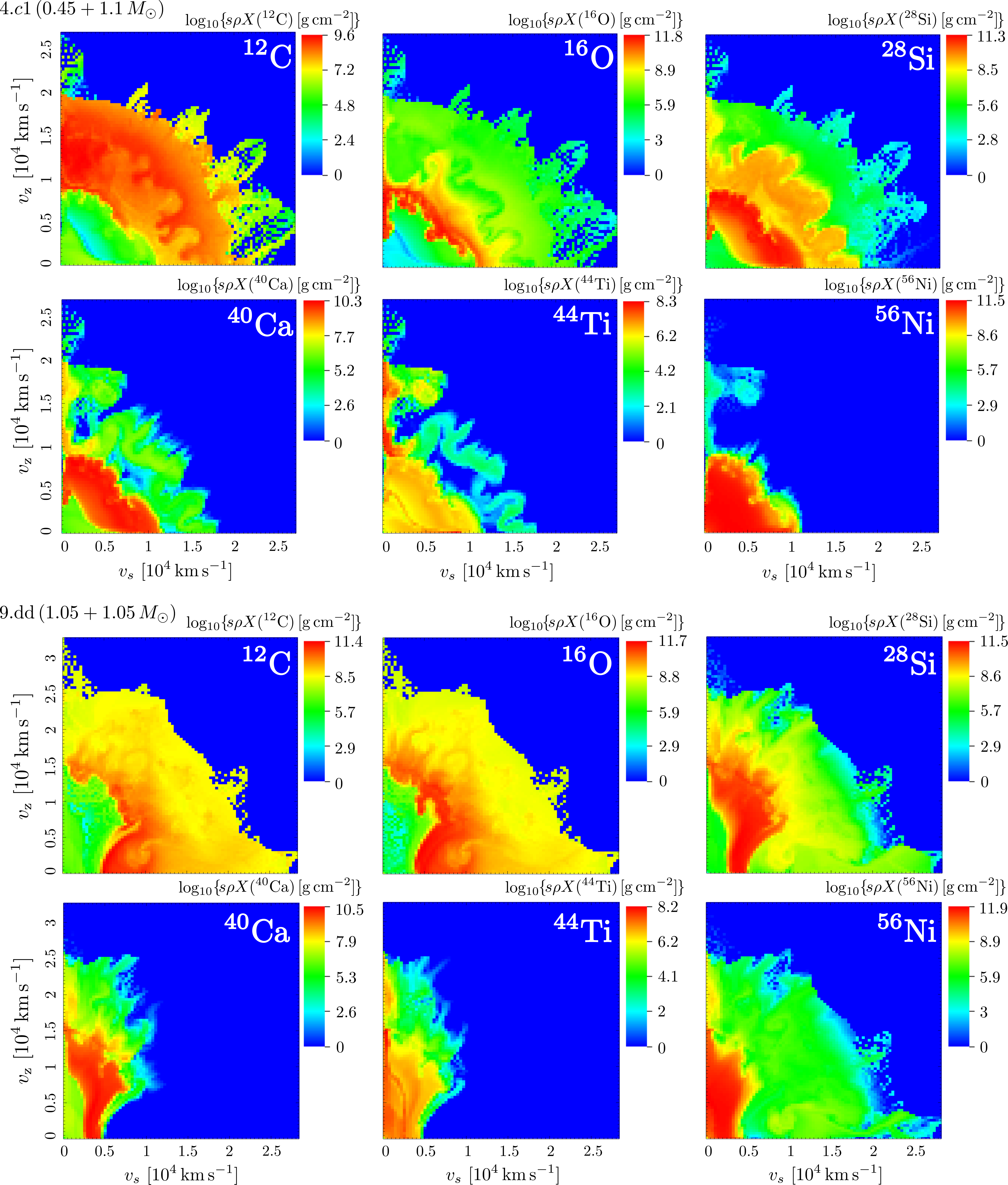}
}
\caption{Final at ($t=39\,{\rm s}$) nuclear mass fractions ($s\rho X_i$, where $s$ is the
  cylindrical radius {\bf [in cm]}, 
  $\rho$ is the density [in ${\rm g\, cm^{-3}}$] and $X_i$ is the mass fraction of the $i$'th
  species; the color scale is logarithmic.)   
  of He, C, O, Si, Ca, Ti and Ni for run 4.c1 ($0.45+1.1\,M_\odot$; upper panels) and 9.dd
  ($1.05+1.05\,M_\odot$; lower panels).}   
\label{fig:isotwosyst}
\end{figure*}

The lower panels of Figure \ref{fig:isotwosyst} show the final nuclear mass fractions
of $\!\,^{12}{\rm C}$, $\!\,^{16}{\rm O}$, $\!\,^{28}{\rm Si}$ and $\!\,^{56}{\rm Ni}$
in the velocity space for the $1.05+1.05\,M_\odot$ system. While the
outer ejecta (i.e., unburned matter) are more or less symmetrically
propagating into the ambient medium, the IGEs have higher ejecta
velocities in the polar direction. 
The most extended shock front (i.e., contact surface between
the ejecta and the low-density ambient matter) is obtained for the
most energetic explosions (Figure \ref{fig:remnants}). The interaction of the
ejecta with the low density ambient gives rise to
Rayleigh-Taylor hydrodynamic instabilities behind the shock front, when
the denser layers penetrate the overlying lighter ambient environment.

\begin{center}
\begin{small}
\begin{table*}
\begin{tabular}{|c|c|c|c|c|c|c|c|}
  \hline\hline
  \multirow{2}{*}{Run} & Temp. & \multirow{2}{*}{$T_{\rm max,8}$} & \multirow{2}{*}{$\rho_5(T_{\rm max})$} 
  & \multicolumn{2}{|c|}{Ignition} & \multirow{2}{*}{$t_{\rm det.env,s}$}
  & \multirow{2}{*}{$t_{\rm
    det.core,s}$} \\*
  \cline{5-6}
   & profile & &  & 
              Envelope &  Core & & \\
\hline\hline
  4.c1-TH  & Top hat & 58.889 & 519.168 & Yes & Yes & 0.1 & 1.54 \\
  4.c1-Linear & Linear & 19.308& 1.950  & Yes &  No & 1.3 & - \\
  4.c1-Gaussian & Gaussian & 59.914 & 655.004 & Yes & Yes & 0.28 & 1.7 \\
  \hline\hline
\end{tabular}
\caption{Dependence on the temperature profiles. The comparison is
  done for the succesful detonating $0.45+1.1\,M_\odot$ system, using
  a perturbation radius $R_{\rm per}$ of 1000 km and the conditions
  returned by the criteria $C_{1}$. The ambient temperature
  corresponds to the maximum temperature on the grid cells surrounding
  the hotspot.}  
\label{tab:Tprof}
\end{table*}
\end{small}
\end{center}

Run 2.dd ($0.45+0.6\,M_\odot$), 3.dd ($0.45+0.9\,M_\odot$) and run
7.dd ($0.6+1.1\,M_\odot$) produce 
little $\,\!^{56}{\rm Ni}$, between 0.01 and $0.09\, M_\odot$, and they would be
faint and lie outside the range of ``normal'' SNe Ia.  
Run 7.dd is the only model where a second detonation in the
(ONe) core could not be triggered (maximum $R_{\rm per} = 1000$ km).

For the $0.6+0.6\,M_\odot$ system, the initial shock wave triggers a
direct detonation at the core's edge (outer regions of 
the core are made from a mixture of CO and He). Very little
$\,\!^{56}{\rm Ni}$ is produced and will thus produce a very faint
supernova, powered mainly by the decay of $\,\!^{48}{\rm Cr}$. The composition
of the ejecta is similar to the model 8HBC1 of \cite{woosley11}, although their 
model is a result from a He detonation. Their model give rise to a dim
fast evolving event (peak magnitude in B-band of -13.5) 
resembling the spectra of sub-luminous SNe, such as SN 1991bg.

Our best candidates to reproduce a ``normal'' 
SN Ia are the ${0.45+1.1\,M_\odot}$ (run 4.c1, discussed above) and
the ${1.05+1.05\,M_\odot}$ (run 9.dd) systems. For the
$1.05+1.05\,M_\odot$ system, the isotope yields and the kinetic energy of the 
ejecta are very similar to those of \cite{moll13} 
for the system with $0.81+0.96\,M_\odot$. The difference in the IMEs
mass is of less than 1\% and in the IGEs of less than 10\%, with their
calculations leading to a larger amount of $\,\!^{56}{\rm Ni}$,
($\sim\! 7\%$ by mass). Our model leaves more unburned material
($50\%$ more C and $30\%$ more O), as the total mass is larger in our
case \citep[$2.1$ vs $1.77\,M_\odot$ in][]{moll13}. However, the
ejecta morphology is very different from \cite{moll13}, where the
detonation is triggered at the merger moment, when the secondary star
has not been disrupted yet.  
The main difference is that in the model of \cite{moll13} the material 
located at the center of the supernova remnant is coming from the
former secondary, relatively low density WD. While there is a high
abundance of IMEs, no $\,\!^{56}{\rm Ni}$ is being produced. In
our model the ejecta has a SN Ia-like structure \citep{mazzali07}, with IGEs
(mainly $\,\!^{56}{\rm Ni}$) at the center surrounded by IMEs and
unburned material in the outer parts of the ejecta.
This nucleosynthesis of this model also compares well with the
$0.9+1.1\,M_\odot$ system from \cite{pakmor12}, just that, again, the
ejecta morphology is different. Their model is more similar to
\cite{moll13}, with the material of the former secondary WD dominating
the central ejecta, and thus no IGEs in the core. 
The more massive ${0.95+1.15\,M_\odot}$ system produces an abundance of
$\,\!^{56}{\rm Ni}$ ($0.82\,M_\odot$) and will most likely produce a
bright SN Ia. As discussed in \cite{moll13}, due to the
asymmetry of the ejecta, the brightness of this model will further be
boosted in the direction where $\,\!^{56}{\rm Ni}$ is closer to the
observer and that could account for the apparent excess mass in
  the ``super-Chandrasekhar'' events. 
A detailed comparison of the nucleosynthetic yields from our models
and those inferred from the observations is presented in
Section \ref{sec:nuc}.

\subsection{Hotspot geometry and composition}

For two systems, with $0.45+0.6\,M_\odot$ and $0.45+0.9\,M_\odot$
components, we have placed the hotspot at the pole. With this setup,
the hotspot has a spherical shape compared to the torus geometry (by
virtue of the cylindrical symmetry) of the other models. 

For the $0.45+0.6\,M_\odot$, run 2.dd-Polar in Table
\ref{tab:results3}, He detonates soon after the start of the 
simulation ($t=0.13$ s) and there is a smooth transition to an
edge-lit CO core detonation at $t\approx 0.19$ s. 
The reduced perturbation radius translates into a reduced perturbation 
energy required to trigger a second detonation in the core by almost
four orders of magnitude. This could be explained by 
the more favorable composition compared to run 2.dd. At the location
of the hotspot, the envelope is not made of
pure He, but a mixture between He ($X[\!\,^{4}{\rm He}]\sim 0.6$),
C ($X[\!\,^{12}{\rm C}]\sim 0.2$) and O ($X[\!\,^{16}{\rm
  O}]\sim 0.2$), while the outer region of the core
where the CO detonation is triggered, the composition is dominated by
C ($X[\!\,^{12}{\rm C}]\sim 0.4$) and O ($X[\!\,^{16}{\rm O}]\sim
0.4$), but mixed with He ($X[\!\,^{4}{\rm He}]\sim 0.4$) (see
\cite{seitenzahl09,shen14} for a discussion on the effects of
composition on the detonation).

For the $0.45+0.9\,M_\odot$ system, run 3.dd-Polar in Table
\ref{tab:results3}, while the perturbation radius has to be larger to
triggere a second detonation in the core compared to run 3.dd, the
deposited energy is almost the same ($5.7$ and $1.6 \times 10^{46}$
for 3.dd and 3.dd-Polar, respectively). 
A He-detonation is triggered after 0.17 s in the upper part of the 
hotspot, where there is more He ($X[\!\,^{4}{\rm He}]\sim 0.7$). It propagates
around the core and it reflects at the $z=0$ symmetry axis after 1.23 s.
Shock waves from the He detonation trigger a second detonation but
only after 2.7 s,  at the symmetry axis $s=0$ and $z \approx 1200$ km
above the core's center.  

The dependence on the temperature profiles of the hotspot has also been
tested. The results of the different setups using a flat, linear and
Gaussian profiles are shown in Table \ref{tab:Tprof}. The different
profiles models follow the same setup as run 4.c1 (see Table
\ref{tab:ICs} in the Appendix), only that here we do not determine
the minimum size $R_{\rm per}$ of the hotspot, but run a single test
for each profile using $R_{\rm per} = 1000$ km. 
The ``top-hat'' profile (run  4.c1-TH) more readily leads to a detonation, shortly
($<0.1$ s) after the simulation are started, the Gaussian
profile (run  4.c1-Gaussian) takes longer to initiate the envelope detonation and
subsequently the core detonation, while for the linear profile (run
4.c1-Linear), a detonation is 
triggered only in the envelope and only after a delay of 1.3 s. The
difference in the evolution of the three models is caused by the
density profiles inside the hotspots. As the density increases from
the center of the hotspot in the direction towards the core, only the
``top-hat'' and Gaussian profiles have a high enough temperature at a
high enough density so that the detonation can quickly emerge.

\subsection{Nucleosynthetic yields}
\label{sec:nuc}

Table \ref{tab:nuc} lists the nucleosynthetic yields of all
successfully detonating models.  Also shown in Table \ref{tab:nuc},
are the final kinetic energies and the sum of the masses of IMEs and
IGEs. Using these and the ejecta composition structure in velocity space,
we constrain our models with a set of observations presented below.
We are using the upper limits placed on the
$\!\,^{44}{\rm  Ti}$ mass from the observations of Tycho's \citep{lopez15}  and
G1.9+1.3's \citep{zoglauer15} remnants, the ejecta velocity of the Si II $\lambda
6355$ and Ca II H\&K features and the total amount of $\!\,^{28}{\rm
  Si}$, $\!\,^{40}{\rm Ca}$ and $\!\,^{56}{\rm Ni}$ synthesised.

To date, there is no direct detection of $\!\,^{44}{\rm Ti}$ in the
SN Ia remnants, but there have been upper limits set on the presence
of $\!\,^{44}{\rm Ti}$. \cite{lopez15}, using NuSTAR observations of Tycho's
remnant, have found an upper limit between 0.47 and $4.1\times
10^{-4}\, M_\odot$, increasing with the distance to the remnant.  
We have also estimated an upper limit for $\!\,^{44}{\rm Ti}$ in
G1.9+1.3, the most recent supernova \citep[most likely a Type
Ia][]{borkowski10,yamaguchi14} in the Milky Way. We used the  
  distance inferred by \cite{reynolds08}  based on an analysis of
  the absorption toward G1.9+0.3, but lower and higher values are
  possible \citep[e.g.,][]{roy14}.
We use the formula presented in \cite{lopez15} to convert flux in the
68 keV to a $\!\,^{44}{\rm Ti}$ mass, where the flux was determined from Figure 6
of \cite{zoglauer15} using a line-width of $1.4\times 10^4\,{\rm
  km\,s^{-1}}$ (or 3 keV at 68 keV) \citep{reynolds08} and a distance
of 8.5 kpc. 

\begin{figure*}
\centerline{
 \includegraphics[height=2.4in]{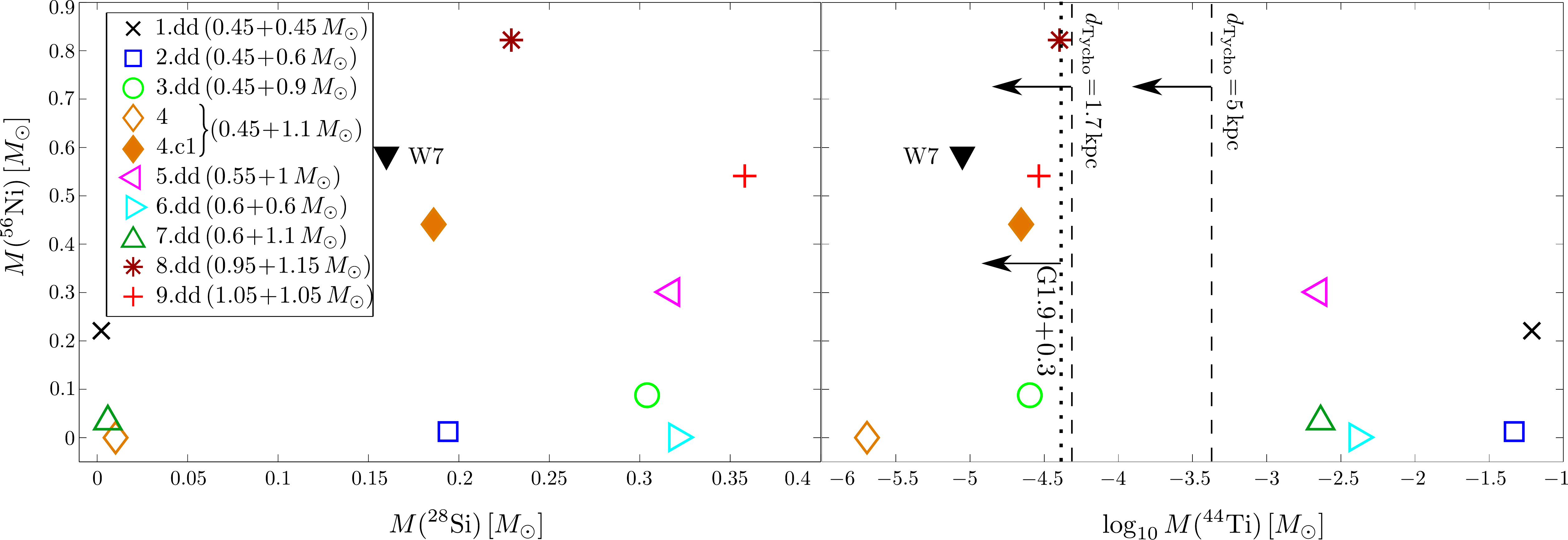}
}
\caption[The synthesised mass of $\,\!^{56}{\rm Ni}$]{The synthesised
  mass of $\,\!^{56}{\rm Ni}$ (in $M_\odot$) as a function of the mass
  of $\,\!^{28}{\rm Si}$ (left) and $\,\!^{44}{\rm Ti}$ (right). The
  three vertical dashed lines in the right panel represent the constraints on
  the $\,\!^{44}{\rm Ti}$ mass for the G1.9+1.3 SN Ia remnant
  \citep[dotted line;][]{zoglauer15} and 
  the Tycho’s SN Ia remnant \citep[dashed line;][]{lopez15}. For Tycho,
  the upper limits of $\,\!^{44}{\rm Ti}$ correspond to range of
  distance estimates in the literature of $d_{\rm Tycho}=1.7-5.0$ 
  kpc. The filled squared marks the W7,
  carbon-deflagration model of \cite{nomoto84b}.  
}  
\label{fig:MNivsMSiMTi}
\end{figure*}

In the right panel of Figure \ref{fig:MNivsMSiMTi}, we plot the 
$\!\,^{56}{\rm Ni}$ mass vs $\!\,^{44}{\rm Ti}$ mass for
all detonating models together with the upper limits from the
observations for the $\!\,^{44}{\rm Ti}$ mass. 
Only five models, located in Figure \ref{fig:MNivsMSiMTi} at the left side with
respect to the vertical lines, meet the constraint on
$M(\!\,^{44}{\rm Ti})$: the two CO mass-transferring systems and the
models involving an initial $0.45$ donor and a $0.9$ and $1.1\,
M_\odot$ accretor. 

A key constraint on the physical conditions in the
SN Ia explosion is the $\!\,^{56}{\rm Ni}$ mass.
The SNe Ia have a large range of luminosities as determined mostly by the
amount of $\!\,^{56}{\rm Ni}$  that is synthesised in the
explosion. The $\!\,^{56}{\rm Ni}$ mass can be determined directly
from the observations assuming that it largely determines the peak
bolometric luminosity \citep{arnett82} and has a key role in
understanding the peak luminosity decline-rate relation
\citep{phillips93}.  
By modelling the late-time nebular spectra of SNe Ia,
\cite{stritzinger06} have accurately measured the $\!\,^{56}{\rm
  Ni}$ mass for a set of 17 SNe Ia, ranging from the sub-luminous
  SN1991bg and up to the bright SN1991T, and found values between
$0.08 $ and $0.94\,M_\odot$. \cite{piro14}, using the volume-limited
sample of 74 SNe Ia from the Lick Observatory Supernova Search
\citep[LOSS;][]{li11a}, found a very similar $\!\,^{56}{\rm Ni}$ range. 
The two closest SN Ia in decades, SN 2011fe \citep{nugent11} and
SN2014J \citep{fossey14}, are also within this range with values varying, depending on
the method used to estimate them. For the SN2011fe,
a $M({\!\,^{56}{\rm Ni}})= 0.47\,M_\odot$ was derived from the nebular spectra
\citep{mazzali15} and a $M({\!\,^{56}{\rm
    Ni}})=0.53\pm0.11\,M_\odot$ from the bolometric luminosity
\citep{pereira13}. For SN 2014J,  $M({\!\,^{56}{\rm Ni}})=0.62\pm
0.13\,M_\odot$ from the gamma-ray lines 
  associated with $\!\,^{56}{\rm Co}$  decay and  from the bolometric
  light curve $M({\!\,^{56}{\rm Ni}})$ within a range of
  $0.57\pm 0.2\,M_\odot$ as a function of  the local extinction due to
  the ambient matter \citep{churazov14}. 

Five models produce $M(\!\,^{56}{\rm Ni})$ between the
observational limits for SNe Ia, as shown in the right panel of Figure 
\ref{fig:MNivsMSiMTi}).  While run 3.dd ($0.45 + 0.9\,M_\odot$) will
probably produce a sub-luminous SN Ia the other
four runs produce  a ${\!\,^{56}{\rm Ni}}$ mass typical for a
``normal'' SN Ia. 

Another constraint on the detonating models comes from the mass of the
IMEs of $\!\,^{28}{\rm Si}$ and $\!\,^{40}{\rm Ca}$. 
In a series of papers, using the code developed by Mazzali and
collaborators, the abundance stratification of five SN
remnants are approximated from a series of late (nebular) spectra
assuming an initial density profile (from W7 and/or a delayed
detonation model).
Values for the $\!\,^{28}{\rm Si}$ and $\!\,^{40}{\rm Ca}$ masses are
generally not given in these works, but they can be extracted
from the ``mass fractions vs. enclosed mass'' figures provided, see
Table \ref{tab:MsiMCa}. 
The values obtained for SN 2002bo, initially by
\cite{stehle05}, have been updated by \cite{blondin15} with a more
accurate model, using non-local 
thermodynamic equilibrium time-dependent radiative-transfer
simulations of a Chandrasekhar-mass delayed-detonation model and we
use their values to constrain the models. 

The $M(\!\,^{28}{\rm Si})$ range from $0.14\, M_\odot$ for SN 1991T
\citep[a peculiar, luminous SN Ia;][]{sasdelli14} up to $0.31\, 
M_\odot$ for SN 2011fe (a ``normal`` SN Ia; \cite{mazzali15}) and
$M(\!\,^{40}{\rm Ca})$ from $8.3\times 10^{-4}\, M_\odot$ for
SN 2011fe up to $0.045\, M_\odot$ for SN 2002bo 
\citep[a ``normal`` SN Ia;][]{blondin15}
In between these ranges are the SN 2003du, a
normal SN Ia \citep{tanaka11} and SN 2004eo, a SN Ia with a luminosity
between normal and sub-luminous SNe Ia \citep{mazzali08}. 
In total, four models do no fit the $M(\!\,^{28}{\rm Si})$ and
$M(\!\,^{40}{\rm Ca})$ range from the observations: single
detonating models, run 4 and 7dd, the double He WD system 1.dd
which is producing too little Si and 2.dd with slightly too much Ca
(see Table \ref{tab:nuc}).

\begin{small}
\begin{table}
\begin{tabular}{|c|c|c|c|}
  \hline\hline
  \multirow{2}{*}{SN} & $M(\!\,^{28}{\rm Si})$ & $M(\!\,^{40}{\rm Ca})$ &
                                                                    \multirow{2}{*}{Reference}
  \\[0.2pt]
  & $[M_\odot]$ & $[M_\odot]$ & \\
  \hline\hline
  \multirow{2}{*}{SN 2002bo} & 0.22 & 0.021 & \cite{stehle05}\\
  & 0.31 & 0.045 & \cite{blondin15}\\
  \hline
  SN 2004eo & 0.41 & $5.7\times 10^{-3}$ & \cite{mazzali08} \\
  \hline
  SN 2003du  & 0.21 & $1.4\times 10^{-3}$ & \cite{tanaka11} \\
  \hline
  SN 1991T & 0.14 & 0.017 & \cite{sasdelli14}) \\
  \hline
  SN 2011fe & 0.31 & $8.3\times 10^{-4}$ & \cite{mazzali15} \\ 
\hline\hline
\end{tabular}
\caption{$\!\,^{28}{\rm Si}$ and $\!\,^{40}{\rm Ca}$ masses for
  five SNe Ia.}
\label{tab:MsiMCa}
\end{table}
\end{small}

\begin{figure*}
\centerline{
\includegraphics[height=8.8in]{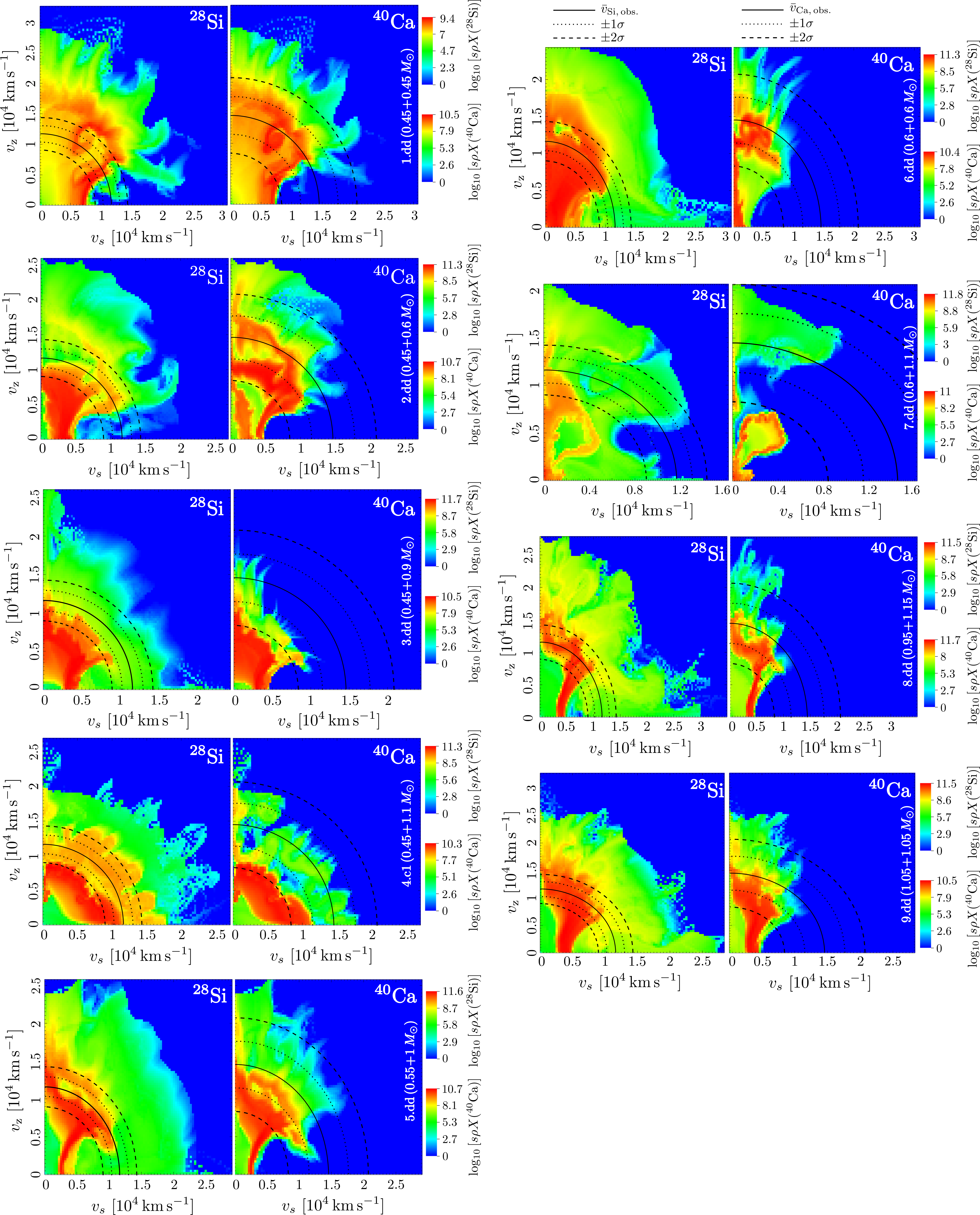}
}
\caption[Si28 and Ca40 velocity]{$\!\,^{28}{\rm
    Si}$ and $\!\,^{40}{\rm Ca}$ ($s \rho X_i$, in
  logarithmic scale; final units are ${\rm g\, cm^{-2}}$, the same as in Figure
  \ref{fig:isotwosyst}, and the snapshots are again taken after
  $t=39\,{\rm s}$) for all remnant models. Overplotted is the  
  mean (continuous line) $1-$ (dotted line) and $2-\sigma$ (dashed
  lines) of the observed Si and Ca velocities near maximum brightness 
  for 91 SNe Ia \citep{foley11a}.}  
\label{fig:vSi_vCa}
\end{figure*}

Si II $\lambda$~6355 and Ca II H\&K are two strong features
in SN Ia spectra. The expansion velocity of
the two features are similar and decrease with time as the ejecta is
expanding and deeper layers of the ejecta are revealed in the
spectra. \cite{foley11a} have estimated the line velocities of Si II
$\lambda$~6355 and Ca\, II\, 3945({H\&K}) near maximum
brightness and we use their 
data for 91 SNe Ia to further constrain the detonating models. 
The SN 2014J and SN 2011fe are also falling within the $2\sigma$ range 
in the sample of \cite{foley11a}, with $v_{\rm Si\, II}=11.9\times
10^3\, {\rm km\,s^{-1}}$ and $v_{\rm Ca\, II}=14.1\times 10^3\, {\rm
  km\,s^{-1}}$ for SN 2014J 
\citep{marion15} and $v_{\rm Si\, II}=10.4\times
10^3\, {\rm km\,s^{-1}}$ for SN 2011fe (no Ca-velocity value 
given; \citep{foley13}. 
Figure \ref{fig:vSi_vCa} shows the $\!\,^{28}{\rm Si}$ and
$\!\,^{40}{\rm Ca}$ yields in velocity space obtained for our
detonating models together with the $1-$ and
$2-\sigma$ lines from the mean of the observed Si and Ca velocities.
While the Si II $\lambda$~6355 velocity near maximum brightness does
not vary significantly amongst the observed SNe Ia (within $\pm
2\sigma$ from the mean, $v_{\rm Si,Obs.}$ is in between $8.9$ and
$14.3\times 10^3\, {\rm km/s}$), Ca II H\&K velocity varies
significantly (between 8.3 and $20.7\times 10^3\, {\rm
  km/s}$, again, within the $2\sigma$ from the mean). 

The mass-weighted mean velocities of Si and Ca of our detonating  
models (see Table \ref{tab:nuc}) lie close to the lower end of the
velocity distribution of the two features, with only 1.dd (both Si and
Ca), 2.dd and 6.dd (only Ca) within 2$\sigma$ of the observations. 
However, the (ionised) mass required to produce a line in
the spectra is very low, with values of $\sim\! 10^{-6}\, M_\odot$ for
Ca II and $\sim\! 10^{-8}\, M_\odot$ for Si II, as computed from
\cite{branch05} using the data provided in their Table 1 for the
optical depths and photosphere velocities corresponding to ten days
after the maximum light and an ion population fraction at the lower
level of transition of 0.05 (a higher number will further decrease the
ion mass required to produce a line). These values can not be
used to further constrain the models, as our models
produce a total Si and Ca mass calculated between the velocity range 
of ``normal'' SNe Ia (i.e., range between $\bar 
v_{\rm Si,Obs}\pm 2\sigma$ and $\bar v_{\rm Ca,Obs}\pm 2\sigma$, see
Figure \ref{fig:vSi_vCa})  between $1.3\times 10^{-5}$ and $4.5\times 
10^{-2} \, M_\odot$ of Ca and between $9.3\times 10^{-4}$ and $0.15 \,
M_\odot$ of Si. If we exclude run 7.dd (single detonating system, with
$0.6+1.1\,M_\odot$ components), the lower limit for Ca mass increases
to $2.4\times 10^{-4}\, M_\odot$.  

\subsection{Initial conditions}
\label{sec:resICs}

Some models trigger the core detonation only when starting with
unrealistic ICs. The model 1.dd ($0.45+0.45\,M_\odot$) has to be
started with a hotspot inside the core in order to trigger its
detonation. This is unrealistic, because, at least when starting with
tidally locked WDs, the core is always colder than
the surrounding region (envelope), even for equal-mass systems
\citep{dan14}. Initial conditions for the 5.dd ($0.55+1\, M_\odot$) and 7.dd
($0.6+1.1\,M_\odot$) models are also not realistic. For these models,
there is very little He ($X(\,\!^{4}{\rm He})<0.1$) in the hot
envelope and, in order to trigger a detonation, the composition inside
the hotspot was artificially set to pure He. 
In some cases, the energy deposition inside the hotspot required to
trigger the core detonation is over a large volume and relatively
high. We label the models 1.dd ($0.45+0.45\,M_\odot$), 2.dd
($0.45+0.6\,M_\odot$) and 8.dd ($0.95+1.15\,M_\odot$) as unrealistic,
as the energy deposition is within an order of magnitude of the
binding energy of the core.

\subsection{Constraints on models}

\begin{center}
\begin{small}
\begin{table*}
  \begin{tabular}{|c|c|c|c|c|c|c|c|c|c|c|c|c|c|}
    \hline\hline
& 1.dd & 2.dd & 3.dd & 4 & 4.c1/c2 & 5.dd  & 6.dd & 7.dd & 8.dd & 9.dd\\
& $0.45\!+\!0.45$ & $0.45\!+\!0.6$ & $0.45\!+\!0.9$ & $0.45\!+\!1.1$ & $0.45\!+\!1.1$ & $0.55\!+\!1$  & $0.6\!+\!0.6$ & $0.6\!+\!1.1$ & $0.95\!+\!1.15$ & $1.05\!+\!1.05$\\
\hline
$\!\,^{28}{\rm Si}$ & \xmark & \cmark & \cmark & \xmark & \cmark & \cmark & \cmark & \xmark & \cmark & \cmark \\[2pt]
$\!\,^{40}{\rm Ca}$ & \cmark & \xmark & \cmark & \xmark & \cmark & \cmark & \cmark & \cmark & \cmark & \cmark \\[2pt]
$\!\,^{44}{\rm Ti}$ & \xmark & \xmark & \cmark & \cmark & \cmark & \xmark & \xmark & \xmark & \cmark & \cmark \\[2pt]
$\!\,^{56}{\rm Ni}$ & \cmark & \xmark & \cmark & \xmark & \cmark & \cmark & \xmark & \xmark & \cmark & \cmark \\[2pt]
ICs & \xmark$^{\rm a,b)}$ & \xmark$^{\rm b)}$ & \cmark & \cmark & \cmark
                                   & \xmark$^{\rm c)}$ & \cmark &
                                                                  \xmark$^{\rm c)}$ & \xmark$^{\rm b)}$  & \cmark  \\[2pt]  
\hline\hline 
\end{tabular}
\caption[Constraints on the models from the observation]{Constraints 
  on the models from the observations of $\!\,^{44}{\rm Ti}$ mass in
  the remnants of Tycho \citep{lopez15} and G1.9+0.3
  \citep{zoglauer15}, $v_{\rm Si}$ and $v_{\rm Ca}$ of ``normal''
  SN Ia sample from \cite{foley11a}, mass of $\!\,^{56}{\rm Ni}$
  of SN Ia ranging from the sub-luminous SN1991bg
  up to the bright SN1991T and from how realistic are the initial
  conditions of each model.\\
  $a)$ hotspot placed artificially inside the (cold) dense core and not within the
  hot envelope which has a low density; \\
  $b)$ $0.45+0.45\,M_\odot$, $0.45+0.6\,M_\odot$ and $0.95+1.15\,M_\odot$
  have a large, unrealistic initial perturbation;\\
  $c)$ because the $0.55+1\, M_\odot$ and $0.6+1.1\,M_\odot$ models have
  very little He inside the hot envelope, the composition inside the hotspot is
  artificially set to pure He. 
} 
\label{tab:constraints}
\end{table*}
\end{small}
\end{center}

We compile Table \ref{tab:constraints} with all the constraints
imposed by the nucleosynthetic yields and initial conditions
presented above. 
By strict application of the criteria only three
models survive: run 3.dd, an 0.45 He with a $0.9 \,M_\odot$ CO WD,
4.c1, an 0.45 He with a $1.1\,M_\odot$ ONe WD and 9.dd, two
$1.05\,M_\odot$ CO WDs. 

For the 4.c1 run, a detonation is ignited in the $1.1\,M_\odot$ ONe
core. While the total mass of the remnant is above the Chandrasekhar
limit, the region between the core and the disk was shock-heated in
the merger and is thermally supported and the (nearly Keplerian)
disk is rotationally supported. While the central density of the ONe
core (initially $\rho_{\rm max}= 3.7\times 10^7\,{\rm g\,cm^{-3}}$) is
increasing as shocks propagate through the core, it never exceeds
$10^8\,{\rm g\,cm^{-3}}$. Thus, the collapse to 
a neutron star can not proceed  as the density threshold
of the electron captures on the $\!\,^{24}{\rm  Mg}$ is above $4\times
10^9\,{\rm  g\,cm^{-3}}$ (in our models $\!\,^{24}{\rm  Mg}$ is
distributed uniformly throughout the core and represents $5\%$ by
mass) and for $\!\,^{20}{\rm  Ne}$ is even higher, at above $9\times
10^9\,{\rm g\,cm^{-3}}$ \citep{miyaji80,saio85}.  
For the 4.c1 model, the shock waves from the initial He detonation
converge inside the core and raise the temperature to $T\approx
6\times 10^9$ K at a density of $\rho\approx 4\times 10^7\, {\rm g\,
  cm^{-3}}$ over a radius of $(100\ {\rm km})$ located at
the pole ($r\approx 0$ km). This triggers a second detonation into the
core. Sufficient energy is released in the nuclear burning
($9.427\times 10^{50}$ erg) to unbind the ONe core (binding energy of
$3.879\times 10^{50}$ erg). 
However, the ONe detonations are not yet proven to work as there are
no calculations of resolved ONe detonation structures (weaker shocks
could yield a detonation in unresolved simulations compared to
resolved ones), with the recent study of \cite{shen14a} arguing against
this possibility due to the increased detonation length- and
shock-strength required to trigger a detonation.  

Recently, \cite{marquardt15} have studied the detonations of massive,
ONe WDs under the assumptions that they ignite at the center of the
star. Compared to our 
models, they start with larger masses, between 1.18 and $1.23\,
M_\odot$, and they are not the remnants of a merger process, but
``naked'', hydrostatic WDs. Their synthetic light curves
do not match the observations of a ``normal'' SN Ia, owing to their
large $\!\,^{56}{\rm Ni}$ masses, but show better agreement with the
``over-luminous'' SN 1991T. Inversely, due to the strong SI II and Ca
II absorption lines shown in the models, the spectra show better
agreement with those of ``normal'' SN Ia than with those of SN 1991T. 
While we can not compare directly to their calculations, we point out
that it is encouraging to note that in our 4.c1 model there is less
$\!\,^{56}{\rm Ni}$ produced, within the range of SNe Ia.

\begin{figure*}
\centerline{
 \includegraphics[height=6.0in]{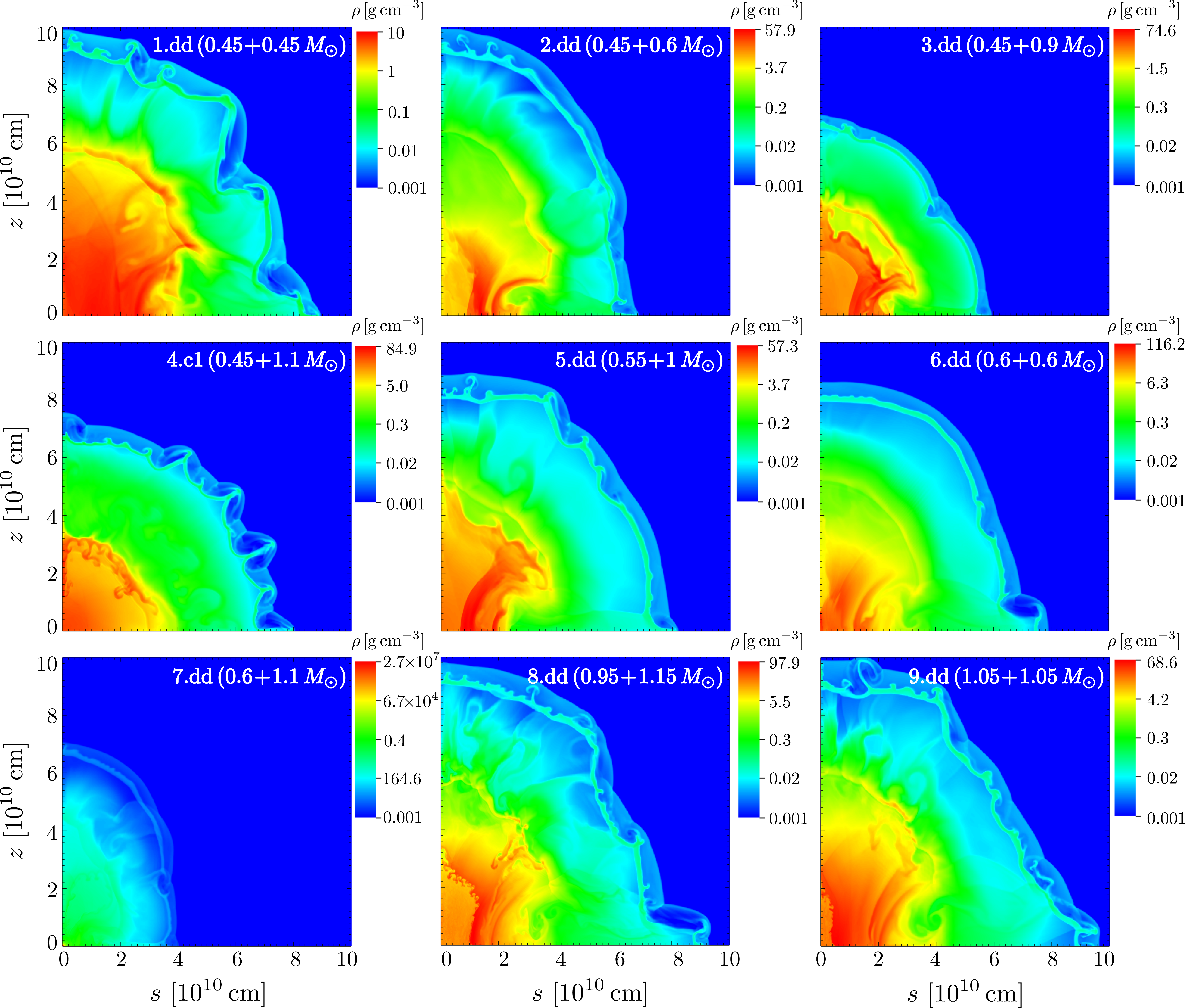}
}
\caption{Density structure of all remnants. 
  The energetics of the ejecta is imprinted 
  in the extension of the shock front at the contact surface between
  the ejecta and the low-density ambient matter.  This is why we had
  to take the final snapshots after $t=39$ s from the
  beginning of the simulations, when for the runs with largest kinetic
  energy the ejecta reach the grid boundaries at $10^{11}$ km.
  For the models shown here a core detonation
  occurs for all but the $0.6+1.1\,M_\odot$ system.   
  The ejecta of the explosion has to propagate through an environment
  spanning a wide range of densities: from the 
  dense, nearly Keplerian-rotating disk to the low density of the ambient. 
  This is naturally resulting in an asymmetric 
  shock front, more extended towards the pole and in developing
  Rayleigh-Taylor hydrodynamic instabilities behind the shock front, when
  the denser layers penetrate the overlying lighter ambient environment. 
}  
\label{fig:remnants}
\end{figure*}

\begin{figure*}
\centerline{
 \includegraphics[height=6.1in]{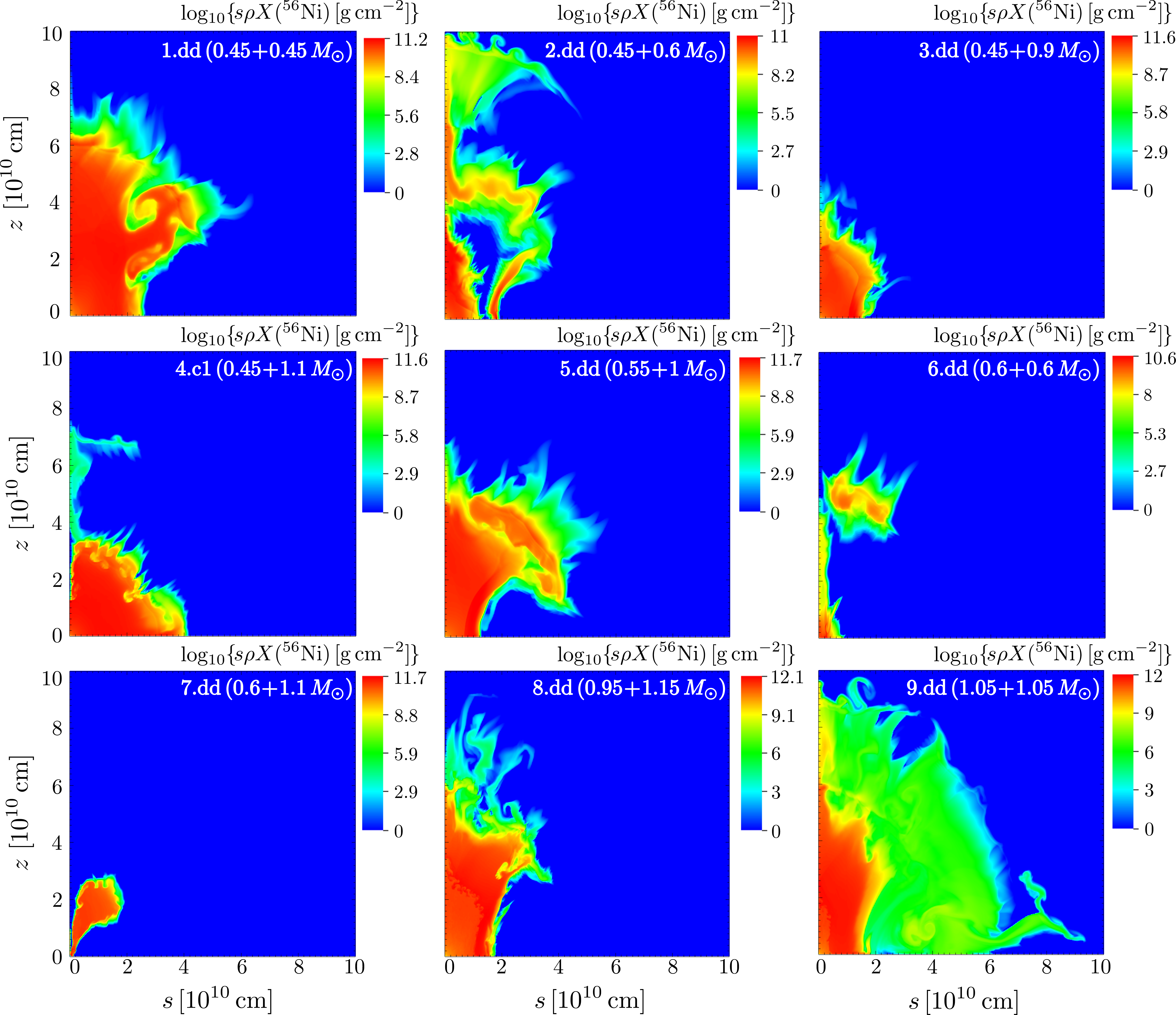}
}
\caption[$\,\!^{56}{\rm Ni}$ mass fraction distribution of all
remnants. ]{$\,\!^{56}{\rm Ni}$ mass fraction distribution (scaled
  with $s\rho$; in logarithmic scale; units are the same as in Figure
  \ref{fig:isotwosyst}) of all remnants.
  Similar to the density structure of the remnants (Figure
  \ref{fig:remnants}) the snapshots were taken after 
  $t=39$ s from the beginning of the simulations and the boundary of
  the domain is at $10^{11}$ km.
  As seen in the density structure, the asymmetric 
  shock front causes a more extended $\,\!^{56}{\rm Ni}$ distribution
  towards the pole.
}  
\label{fig:Ni56}
\end{figure*}

There are strong orientation effects for most of the models as the nuclear
burning products are moving faster in the polar direction (Figure
\ref{fig:remnants} and \ref{fig:Ni56}). The
asymmetries are caused by the location where the detonation is triggered in the 
envelope/core, the slowing of the ejecta by the surrounding
disk, concentrated in the equatorial plane and by the rotation. The
light curves resulting from these models would likely show a non-negligible
sensitivity to line-of-sight effects \citep[see a detailed discussion on the
line-of-sight effects for asymmetric ejecta in][]{moll13}.

\section{Summary}
\label{sec:summary}

We have studied 2D detonation models of WD-WD merger remnants using
the Eulerian AMR code FLASH. Our initial conditions here are based on
the remnant structures from our earlier 3D SPH simulations \citep{dan14}. 
In total we consider nine systems that cover the entire range of WD
masses and compositions.  For each of these systems, we modeled the
initiation and  propagation of detonations and followed the
nucleosynthesis processes until the ejecta reached the homologous
phase. 

After the restart of the simulations with FLASH only the $0.45+1.1\,M_\odot$
model (run 4) triggered a detonation in the hot envelope, but it did not lead
to a second detonation in the underlying core. The envelope detonation
produced almost no $\!\,^{56}{\rm Ni}$ and very little other
radioactive elements. 
Note, however, the results for this set of runs should be regarded as
lower limits.
There are two main reasons that these models do not detonate. First,
the 3D SPH calculations were done at a moderate resolution and, in
reality, mergers will produce remnants that are more prone to
explosion than our finite resolution results. Secondly, there is a
loss of resolution after the mapping onto the grid. This has an impact
onto the thermal evolution of the remnant as the hotspots in the hot
envelope surrounding the relatively cold core are smoothed out. Chemical
compositions are also deteriorated by the mapping process, especially
inside chemically mixed regions. 
This motivated two
more sets of runs, with different setups for the initiation of a
detonation being tested. In one set of models, we manually setup
hotspots in the remnant's envelopes based on the realistic
conditions found in the 3D SPH simulations. In another set of models,
the hotspots were setup based on the conditions for direct detonation
initiation taken from the results of spatially resolved 1D
calculations from the literature. It is in this
later set of models that all systems trigger a detonation, whether it
is only an envelope detonation or the core is ignited as well. 

With perturbations based on the several SPH criteria, again only
the $0.45+1.1\,M_\odot$ system detonates, but this time the envelope
detonation is strong enough so that a second detonation occurs in the
ONe core. The model which leads to a double detonation, run 4.c1, starts with
a hotspot at the location where the ratio $\tau_{\rm 
  nuc}/\tau_{\rm dyn}(T)$ is minimum. The hotspot's temperature is the
same with the temperature of the SPH particle with minimum
ratio.

With perturbations following the critical conditions for a direct 
initiation of a detonation from the spatially resolved 1D
calculations, all systems detonate. Three possible outcomes were
found: a detonation is triggered in the envelope but a second
detonation in the core is avoided (run 7.dd); a detonation is not
ignited in the envelope but the shock waves from the initial
perturbation converge inside the core and trigger a detonation there
(run 6.dd and runs with CO mass transferring systems, 8.dd and 9.dd);
for the other models a detonations occur, both, in the He shell and the
CO/ONe core. However, in some cases the initial conditions are
unrealistic and we reject these models as viable SNe Ia progenitor
candidates.  

The nucleosynthetic yields of the successfully detonating models 
have been compared with the observations of SNe Ia and their
subsequent constraints have been applied to the WD-WD merger
scenario. 
Only three models survive all constraints and potentially
lead to a SN Ia event: $0.45\, M_\odot$ He and $0.9 \,M_\odot$ CO WD that
produces little $\!\,^{56}{\rm Ni}$ and could possibly result in a
sub-luminous, SN 1991bg-like event and two  
good candidates for reproducing common SNe Ia, $0.45\ M_\odot$ He with
a $1.1\,M_\odot$ ONe WD and a double CO WD system with two
$1.05\,M_\odot$ components.
The last two models have in common a former accretor with a mass
$M>1\,M_\odot$, while the last model has a total mass well above the
Chandrasekhar limit. There is observational evidence that such massive WDs
exist and they form a mass excess in the mass distribution
\citep[e.g.,][]{liebert05,ferrario05}. Population synthesis
calculations of \cite{fryer10} also predict that over one third of the
systems with a total mass above the Chandrasekhar mass limit have
$M_{\rm tot}\gtrsim 1.8\,M_\odot$. However, they do not dominate the
current SN Ia sample being less common than systems with a combined
mass near the Chandrasekhar limit \citep{toonen12}, such as the
$0.45+0.9 \,M_\odot$ system.   
Our results confirm that WD-WD mergers can show reasonable  
agreement with the observed nucleosynthesis of SNe Ia and a variety of
outcomes are possible, covering essentially the entire range of
observed supernovae that are associated with thermonuclear
explosions.

\section*{Acknowledgments}
We thank the reviewer for his/her valuable comments and suggestions
that improved the manuscript. 
We thank Laura Lopez, Jerod Parrent, Danny Milisavljevic, Lev
Yungelson and Philipp Podsiadlowski for discussions and for
suggesting useful references. This work was supported by
  Einstein grant PF3-140108 (J. G.), the Packard grant (E. R.), NASA ATP
 grant NNX14AH37G (E. R.). and the Swedish 
 Research Council (VR) under grant 621-2012-4870 (S. R.). 
The FLASH code was developed in part by the DOE-supported Alliances
Center for Astrophysical Thermonuclear Flashes (ASC) at the
University of Chicago. 

\bibliographystyle{mn2e}
\bibliography{flash_det_v3}

\begin{thebibliography}{96}
\expandafter\ifx\csname natexlab\endcsname\relax\def\natexlab#1{#1}\fi

\bibitem[{{Arnett}(1982)}]{arnett82}
{Arnett} W.~D., 1982, ApJ, 253, 785

\bibitem[{{Aznar-Sigu{\'a}n} {et~al}\mbox{.}(2013){Aznar-Sigu{\'a}n},
  {Garc{\'{\i}}a-Berro}, {Lor{\'e}n-Aguilar}, {Jos{\'e}}, \& {Isern}}]{aznar13}
{Aznar-Sigu{\'a}n} G., {Garc{\'{\i}}a-Berro} E., {Lor{\'e}n-Aguilar} P.,
  {Jos{\'e}} J., {Isern} J., 2013, MNRAS, 434, 2539

\bibitem[{{Bildsten} {et~al}\mbox{.}(2007){Bildsten}, {Shen}, {Weinberg}, \&
  {Nelemans}}]{bildsten07}
{Bildsten} L., {Shen} K.~J., {Weinberg} N.~N., {Nelemans} G., 2007, ApJL, 662,
  L95

\bibitem[{{Blondin} {et~al}\mbox{.}(2015){Blondin}, {Dessart}, \&
  {Hillier}}]{blondin15}
{Blondin} S., {Dessart} L., {Hillier} D.~J., 2015, MNRAS, 448, 2766

\bibitem[{{Borkowski} {et~al}\mbox{.}(2010){Borkowski}, {Reynolds}, {Green},
  {Hwang}, {Petre}, {Krishnamurthy}, \& {Willett}}]{borkowski10}
{Borkowski} K.~J., {Reynolds} S.~P., {Green} D.~A., {Hwang} U., {Petre} R.,
  {Krishnamurthy} K., {Willett} R., 2010, ApJL, 724, L161

\bibitem[{{Branch} {et~al}\mbox{.}(2005){Branch}, {Baron}, {Hall}, {Melakayil},
  \& {Parrent}}]{branch05}
{Branch} D., {Baron} E., {Hall} N., {Melakayil} M., {Parrent} J., 2005, PASP,
  117, 545

\bibitem[{{Churazov} {et~al}\mbox{.}(2014){Churazov}, {Sunyaev}, {Isern},
  {Kn{\"o}dlseder}, {Jean}, {Lebrun}, {Chugai}, {Grebenev}, {Bravo}, {Sazonov},
  \& {Renaud}}]{churazov14}
{Churazov} E. {et~al.}, 2014, Nature, 512, 406

\bibitem[{{Couch} {et~al}\mbox{.}(2013){Couch}, {Graziani}, \&
  {Flocke}}]{couch13}
{Couch} S.~M., {Graziani} C., {Flocke} N., 2013, ApJ, 778, 181

\bibitem[{{Dan} {et~al}\mbox{.}(2014){Dan}, {Rosswog}, {Br{\"u}ggen}, \&
  {Podsiadlowski}}]{dan14}
{Dan} M., {Rosswog} S., {Br{\"u}ggen} M., {Podsiadlowski} P., 2014, MNRAS, 438,
  14

\bibitem[{{Dan} {et~al}\mbox{.}(2012){Dan}, {Rosswog}, {Guillochon}, \&
  {Ramirez-Ruiz}}]{dan12}
{Dan} M., {Rosswog} S., {Guillochon} J., {Ramirez-Ruiz} E., 2012, MNRAS, 422,
  2417

\bibitem[{{Ferrario} {et~al}\mbox{.}(2005){Ferrario}, {Wickramasinghe},
  {Liebert}, \& {Williams}}]{ferrario05}
{Ferrario} L., {Wickramasinghe} D., {Liebert} J., {Williams} K.~A., 2005,
  MNRAS, 361, 1131

\bibitem[{{Fink} {et~al}\mbox{.}(2007){Fink}, {Hillebrandt}, \&
  {R{\"o}pke}}]{fink07}
{Fink} M., {Hillebrandt} W., {R{\"o}pke} F.~K., 2007, A\&A, 476, 1133

\bibitem[{{Foley} \& {Kirshner}(2013)}]{foley13}
{Foley} R.~J., {Kirshner} R.~P., 2013, ApJL, 769, L1

\bibitem[{{Foley} {et~al}\mbox{.}(2011){Foley}, {Sanders}, \&
  {Kirshner}}]{foley11a}
{Foley} R.~J., {Sanders} N.~E., {Kirshner} R.~P., 2011, ApJ, 742, 89

\bibitem[{{Fossey} {et~al}\mbox{.}(2014){Fossey}, {Cooke}, {Pollack}, {Wilde},
  \& {Wright}}]{fossey14}
{Fossey} J., {Cooke} B., {Pollack} G., {Wilde} M., {Wright} T., 2014, CBET,
  3792, 1

\bibitem[{{Fryer} {et~al}\mbox{.}(2010){Fryer}, {Ruiter}, {Belczynski},
  {Brown}, {Bufano}, {Diehl}, {Fontes}, {Frey}, {Holland}, {Hungerford},
  {Immler}, {Mazzali}, {Meakin}, {Milne}, {Raskin}, \& {Timmes}}]{fryer10}
{Fryer} C.~L. {et~al.}, 2010, ApJ, 725, 296

\bibitem[{{Fryxell} {et~al}\mbox{.}(2000){Fryxell}, {Olson}, {Ricker},
  {Timmes}, {Zingale}, {Lamb}, {MacNeice}, {Rosner}, {Truran}, \&
  {Tufo}}]{fryxell00}
{Fryxell} B. {et~al.}, 2000, ApJS, 131, 273

\bibitem[{{Fryxell} {et~al}\mbox{.}(1989){Fryxell}, {M\" uller}, \&
  {Arnett}}]{fryxell89}
{Fryxell} B.~A., {M\" uller} E., {Arnett} W.~D., 1989, MPA Green Report,
  Max-Planck-Institut für Astrophysik Garching, 449

\bibitem[{{Garc{\'{\i}}a-Senz} {et~al}\mbox{.}(1999){Garc{\'{\i}}a-Senz},
  {Bravo}, \& {Woosley}}]{garcia99}
{Garc{\'{\i}}a-Senz} D., {Bravo} E., {Woosley} S.~E., 1999, A\&A, 349, 177

\bibitem[{{Garc{\'{\i}}a-Senz} {et~al}\mbox{.}(2013){Garc{\'{\i}}a-Senz},
  {Cabez{\'o}n}, {Arcones}, {Rela{\~n}o}, \& {Thielemann}}]{garcia13}
{Garc{\'{\i}}a-Senz} D., {Cabez{\'o}n} R.~M., {Arcones} A., {Rela{\~n}o} A.,
  {Thielemann} F.~K., 2013, MNRAS, 436, 3413

\bibitem[{{Guerrero} {et~al}\mbox{.}(2004){Guerrero}, {Garc{\'{\i}}a-Berro}, \&
  {Isern}}]{guerrero04}
{Guerrero} J., {Garc{\'{\i}}a-Berro} E., {Isern} J., 2004, A\&A, 413, 257

\bibitem[{{Guillochon} {et~al}\mbox{.}(2010){Guillochon}, {Dan},
  {Ramirez-Ruiz}, \& {Rosswog}}]{guillochon10}
{Guillochon} J., {Dan} M., {Ramirez-Ruiz} E., {Rosswog} S., 2010, ApJL, 709,
  L64

\bibitem[{{Hawley} {et~al}\mbox{.}(2012){Hawley}, {Athanassiadou}, \&
  {Timmes}}]{hawley12}
{Hawley} W.~P., {Athanassiadou} T., {Timmes} F.~X., 2012, ApJ, 759, 39

\bibitem[{{Hillebrandt} \& {Niemeyer}(2000)}]{hillebrandt00}
{Hillebrandt} W., {Niemeyer} J.~C., 2000, ARAA, 38, 191

\bibitem[{{Holcomb} {et~al}\mbox{.}(2013){Holcomb}, {Guillochon}, {De Colle},
  \& {Ramirez-Ruiz}}]{holcomb13}
{Holcomb} C., {Guillochon} J., {De Colle} F., {Ramirez-Ruiz} E., 2013, ApJ,
  771, 14

\bibitem[{{Howell}(2011)}]{howell11}
{Howell} D.~A., 2011, Nature Communications, 2

\bibitem[{{Iben} \& {Tutukov}(1984)}]{iben84}
{Iben}, Jr. I., {Tutukov} A.~V., 1984, ApJS, 54, 335

\bibitem[{{Kashi} \& {Soker}(2011)}]{kashi11}
{Kashi} A., {Soker} N., 2011, MNRAS, 417, 1466

\bibitem[{{Kashyap} {et~al}\mbox{.}(2015){Kashyap}, {Fisher},
  {Garc{\'{\i}}a-Berro}, {Aznar-Sigu{\'a}n}, {Ji}, \&
  {Lor{\'e}n-Aguilar}}]{kashyap15}
{Kashyap} R., {Fisher} R., {Garc{\'{\i}}a-Berro} E., {Aznar-Sigu{\'a}n} G.,
  {Ji} S., {Lor{\'e}n-Aguilar} P., 2015, ApJL, 800, L7

\bibitem[{{Kushnir} {et~al}\mbox{.}(2013){Kushnir}, {Katz}, {Dong}, {Livne}, \&
  {Fern{\'a}ndez}}]{kushnir13}
{Kushnir} D., {Katz} B., {Dong} S., {Livne} E., {Fern{\'a}ndez} R., 2013, ApJL,
  778, L37

\bibitem[{{Li} {et~al}\mbox{.}(2011){Li}, {Leaman}, {Chornock}, {Filippenko},
  {Poznanski}, {Ganeshalingam}, {Wang}, {Modjaz}, {Jha}, {Foley}, \&
  {Smith}}]{li11a}
{Li} W. {et~al.}, 2011, MNRAS, 412, 1441

\bibitem[{{Liebert} {et~al}\mbox{.}(2005){Liebert}, {Bergeron}, \&
  {Holberg}}]{liebert05}
{Liebert} J., {Bergeron} P., {Holberg} J.~B., 2005, ApJS, 156, 47

\bibitem[{{Livio} \& {Riess}(2003)}]{livio03}
{Livio} M., {Riess} A.~G., 2003, ApJL, 594, L93

\bibitem[{{Livne}(1990)}]{livne90}
{Livne} E., 1990, ApJL, 354, L53

\bibitem[{{Livne} \& {Arnett}(1995)}]{livne95}
{Livne} E., {Arnett} D., 1995, ApJ, 452, 62

\bibitem[{{Lopez} {et~al}\mbox{.}(2015){Lopez}, {Grefenstette}, {Reynolds},
  {An}, {Boggs}, {Christensen}, {Craig}, {Eriksen}, {Fryer}, {Hailey},
  {Harrison}, {Madsen}, {Stern}, {Zhang}, \& {Zoglauer}}]{lopez15}
{Lopez} L.~A. {et~al.}, 2015, ArXiv e-prints, arXiv:1504.07238

\bibitem[{{Luminet} \& {Pichon}(1989)}]{luminet89}
{Luminet} J.-P., {Pichon} B., 1989, A\&A, 209, 103

\bibitem[{MacNeice {et~al}\mbox{.}(2000)MacNeice, Olson, Mobarry,
  de~Fainchtein, \& Packer}]{macneice00}
MacNeice P., Olson K.~M., Mobarry C., de~Fainchtein R., Packer C., 2000,
  Computer Physics Communications, 126, 330

\bibitem[{{Maoz} {et~al}\mbox{.}(2014){Maoz}, {Mannucci}, \&
  {Nelemans}}]{maoz14}
{Maoz} D., {Mannucci} F., {Nelemans} G., 2014, ARA\&A, 52, 107

\bibitem[{{Marion} {et~al}\mbox{.}(2015){Marion}, {Sand}, {Hsiao}, {Banerjee},
  {Valenti}, {Stritzinger}, {Vink{\'o}}, {Joshi}, {Venkataraman}, {Ashok},
  {Amanullah}, {Binzel}, {Bochanski}, {Bryngelson}, {Burns}, {Drozdov},
  {Fieber-Beyer}, {Graham}, {Howell}, {Johansson}, {Kirshner}, {Milne},
  {Parrent}, {Silverman}, {Vervack}, \& {Wheeler}}]{marion15}
{Marion} G.~H. {et~al.}, 2015, ApJ, 798, 39

\bibitem[{{Marquardt} {et~al}\mbox{.}(2015){Marquardt}, {Sim}, {Ruiter},
  {Seitenzahl}, {Ohlmann}, {Kromer}, {Pakmor}, \& {R{\"o}pke}}]{marquardt15}
{Marquardt} K.~S., {Sim} S.~A., {Ruiter} A.~J., {Seitenzahl} I.~R., {Ohlmann}
  S.~T., {Kromer} M., {Pakmor} R., {R{\"o}pke} F.~K., 2015, A\&A, 580, A118

\bibitem[{{Mazzali} {et~al}\mbox{.}(2007){Mazzali}, {R{\"o}pke}, {Benetti}, \&
  {Hillebrandt}}]{mazzali07}
{Mazzali} P.~A., {R{\"o}pke} F.~K., {Benetti} S., {Hillebrandt} W., 2007,
  Science, 315, 825

\bibitem[{{Mazzali} {et~al}\mbox{.}(2008){Mazzali}, {Sauer}, {Pastorello},
  {Benetti}, \& {Hillebrandt}}]{mazzali08}
{Mazzali} P.~A., {Sauer} D.~N., {Pastorello} A., {Benetti} S., {Hillebrandt}
  W., 2008, MNRAS, 386, 1897

\bibitem[{{Mazzali} {et~al}\mbox{.}(2015){Mazzali}, {Sullivan}, {Filippenko},
  {Garnavich}, {Clubb}, {Maguire}, {Pan}, {Shappee}, {Silverman}, {Benetti},
  {Hachinger}, {Nomoto}, \& {Pian}}]{mazzali15}
{Mazzali} P.~A. {et~al.}, 2015, MNRAS, 450, 2631

\bibitem[{{Miyaji} {et~al}\mbox{.}(1980){Miyaji}, {Nomoto}, {Yokoi}, \&
  {Sugimoto}}]{miyaji80}
{Miyaji} S., {Nomoto} K., {Yokoi} K., {Sugimoto} D., 1980, PASJ, 32, 303

\bibitem[{{Moll} \& {Woosley}(2013)}]{moll13}
{Moll} R., {Woosley} S.~E., 2013, ApJ, 774, 137

\bibitem[{{Nomoto}(1982{\natexlab{a}})}]{nomoto82b}
{Nomoto} K., 1982{\natexlab{a}}, ApJ, 257, 780

\bibitem[{{Nomoto}(1982{\natexlab{b}})}]{nomoto82a}
{Nomoto} K., 1982{\natexlab{b}}, ApJ, 253, 798

\bibitem[{{Nomoto} {et~al}\mbox{.}(1984){Nomoto}, {Thielemann}, \&
  {Yokoi}}]{nomoto84b}
{Nomoto} K., {Thielemann} F.-K., {Yokoi} K., 1984, ApJ, 286, 644

\bibitem[{{Nugent} {et~al}\mbox{.}(2011){Nugent}, {Sullivan}, {Cenko},
  {Thomas}, {Kasen}, {Howell}, {Bersier}, {Bloom}, {Kulkarni}, {Kandrashoff},
  {Filippenko}, {Silverman}, {Marcy}, {Howard}, {Isaacson}, {Maguire},
  {Suzuki}, {Tarlton}, {Pan}, {Bildsten}, {Fulton}, {Parrent}, {Sand},
  {Podsiadlowski}, {Bianco}, {Dilday}, {Graham}, {Lyman}, {James}, {Kasliwal},
  {Law}, {Quimby}, {Hook}, {Walker}, {Mazzali}, {Pian}, {Ofek}, {Gal-Yam}, \&
  {Poznanski}}]{nugent11}
{Nugent} P.~E. {et~al.}, 2011, Nature, 480, 344

\bibitem[{{Pakmor} {et~al}\mbox{.}(2010){Pakmor}, {Kromer}, {R{\"o}pke}, {Sim},
  {Ruiter}, \& {Hillebrandt}}]{pakmor10}
{Pakmor} R., {Kromer} M., {R{\"o}pke} F.~K., {Sim} S.~A., {Ruiter} A.~J.,
  {Hillebrandt} W., 2010, Nature, 463, 61

\bibitem[{{Pakmor} {et~al}\mbox{.}(2012){Pakmor}, {Kromer}, {Taubenberger},
  {Sim}, {R{\"o}pke}, \& {Hillebrandt}}]{pakmor12}
{Pakmor} R., {Kromer} M., {Taubenberger} S., {Sim} S.~A., {R{\"o}pke} F.~K.,
  {Hillebrandt} W., 2012, ApJL, 747, L10

\bibitem[{{Pakmor} {et~al}\mbox{.}(2013){Pakmor}, {Kromer}, {Taubenberger}, \&
  {Springel}}]{pakmor13}
{Pakmor} R., {Kromer} M., {Taubenberger} S., {Springel} V., 2013, ApJL, 770, L8

\bibitem[{{Papatheodore} \& {Messer}(2014)}]{papatheodore14}
{Papatheodore} T.~L., {Messer} O.~E.~B., 2014, ApJ, 782, 12

\bibitem[{{Papish} \& {Perets}(2015)}]{papish15}
{Papish} O., {Perets} H.~B., 2015, ArXiv e-prints, arXiv:1502.03453

\bibitem[{{Pereira} {et~al}\mbox{.}(2013){Pereira}, {Thomas}, {Aldering},
  {Antilogus}, {Baltay}, {Benitez-Herrera}, {Bongard}, {Buton}, {Canto},
  {Cellier-Holzem}, {Chen}, {Childress}, {Chotard}, {Copin}, {Fakhouri},
  {Fink}, {Fouchez}, {Gangler}, {Guy}, {Hillebrandt}, {Hsiao}, {Kerschhaggl},
  {Kowalski}, {Kromer}, {Nordin}, {Nugent}, {Paech}, {Pain}, {P{\'e}contal},
  {Perlmutter}, {Rabinowitz}, {Rigault}, {Runge}, {Saunders}, {Smadja}, {Tao},
  {Taubenberger}, {Tilquin}, \& {Wu}}]{pereira13}
{Pereira} R. {et~al.}, 2013, A\&A, 554, A27

\bibitem[{{Perlmutter} {et~al}\mbox{.}(1998){Perlmutter}, {Aldering}, {della
  Valle}, {Deustua}, {Ellis}, {Fabbro}, {Fruchter}, {Goldhaber}, {Groom},
  {Hook}, {Kim}, {Kim}, {Knop}, {Lidman}, {McMahon}, {Nugent}, {Pain},
  {Panagia}, {Pennypacker}, {Ruiz-Lapuente}, {Schaefer}, \&
  {Walton}}]{perlmutter98}
{Perlmutter} S. {et~al.}, 1998, Nature, 391, 51

\bibitem[{{Phillips}(1993)}]{phillips93}
{Phillips} M.~M., 1993, ApJL, 413, L105

\bibitem[{{Piro} {et~al}\mbox{.}(2014){Piro}, {Thompson}, \&
  {Kochanek}}]{piro14}
{Piro} A.~L., {Thompson} T.~A., {Kochanek} C.~S., 2014, MNRAS, 438, 3456

\bibitem[{{Postnov} \& {Yungelson}(2014)}]{postnov14}
{Postnov} K.~A., {Yungelson} L.~R., 2014, Living Rev. in Relativ., 17, 3

\bibitem[{{Raskin} {et~al}\mbox{.}(2014){Raskin}, {Kasen}, {Moll}, {Schwab}, \&
  {Woosley}}]{raskin14}
{Raskin} C., {Kasen} D., {Moll} R., {Schwab} J., {Woosley} S., 2014, ApJ, 788,
  75

\bibitem[{{Raskin} {et~al}\mbox{.}(2012){Raskin}, {Scannapieco}, {Fryer},
  {Rockefeller}, \& {Timmes}}]{raskin12}
{Raskin} C., {Scannapieco} E., {Fryer} C., {Rockefeller} G., {Timmes} F.~X.,
  2012, ApJ, 746, 62

\bibitem[{{Raskin} {et~al}\mbox{.}(2009){Raskin}, {Timmes}, {Scannapieco},
  {Diehl}, \& {Fryer}}]{raskin09}
{Raskin} C., {Timmes} F.~X., {Scannapieco} E., {Diehl} S., {Fryer} C., 2009,
  MNRAS, 399, L156

\bibitem[{{Reynolds} {et~al}\mbox{.}(2008){Reynolds}, {Borkowski}, {Green},
  {Hwang}, {Harrus}, \& {Petre}}]{reynolds08}
{Reynolds} S.~P., {Borkowski} K.~J., {Green} D.~A., {Hwang} U., {Harrus} I.,
  {Petre} R., 2008, ApJL, 680, L41

\bibitem[{{Riess} {et~al}\mbox{.}(1998){Riess}, {Filippenko}, {Challis},
  {Clocchiatti}, {Diercks}, {Garnavich}, {Gilliland}, {Hogan}, {Jha},
  {Kirshner}, {Leibundgut}, {Phillips}, {Reiss}, {Schmidt}, {Schommer},
  {Smith}, {Spyromilio}, {Stubbs}, {Suntzeff}, \& {Tonry}}]{riess98}
{Riess} A.~G. {et~al.}, 1998, AJ, 116, 1009

\bibitem[{{R{\"o}pke} {et~al}\mbox{.}(2007){R{\"o}pke}, {Woosley}, \&
  {Hillebrandt}}]{roepke07}
{R{\"o}pke} F.~K., {Woosley} S.~E., {Hillebrandt} W., 2007, ApJ, 660, 1344

\bibitem[{{Rosswog} {et~al}\mbox{.}(2009{\natexlab{a}}){Rosswog}, {Kasen},
  {Guillochon}, \& {Ramirez-Ruiz}}]{rosswog09c}
{Rosswog} S., {Kasen} D., {Guillochon} J., {Ramirez-Ruiz} E.,
  2009{\natexlab{a}}, ApJL, 705, L128

\bibitem[{{Rosswog} {et~al}\mbox{.}(2009{\natexlab{b}}){Rosswog},
  {Ramirez-Ruiz}, \& {Hix}}]{rosswog09b}
{Rosswog} S., {Ramirez-Ruiz} E., {Hix} W.~R., 2009{\natexlab{b}}, ApJ, 695, 404

\bibitem[{{Roy} \& {Pal}(2014)}]{roy14}
{Roy} S., {Pal} S., 2014, in IAU Symposium, Vol. 296, IAU Symposium, {Ray} A.,
  {McCray} R.~A., eds., pp. 197--201

\bibitem[{{Saio} \& {Nomoto}(1985)}]{saio85}
{Saio} H., {Nomoto} K., 1985, A\&A, 150, L21

\bibitem[{{Sasdelli} {et~al}\mbox{.}(2014){Sasdelli}, {Mazzali}, {Pian},
  {Nomoto}, {Hachinger}, {Cappellaro}, \& {Benetti}}]{sasdelli14}
{Sasdelli} M., {Mazzali} P.~A., {Pian} E., {Nomoto} K., {Hachinger} S.,
  {Cappellaro} E., {Benetti} S., 2014, MNRAS, 445, 711

\bibitem[{{Sato} {et~al}\mbox{.}(2015){Sato}, {Nakasato}, {Tanikawa}, {Nomoto},
  {Maeda}, \& {Hachisu}}]{sato15}
{Sato} Y., {Nakasato} N., {Tanikawa} A., {Nomoto} K., {Maeda} K., {Hachisu} I.,
  2015, ApJ, 807, 105

\bibitem[{{Seitenzahl} {et~al}\mbox{.}(2009){Seitenzahl}, {Meakin}, {Townsley},
  {Lamb}, \& {Truran}}]{seitenzahl09}
{Seitenzahl} I.~R., {Meakin} C.~A., {Townsley} D.~M., {Lamb} D.~Q., {Truran}
  J.~W., 2009, ApJ, 696, 515

\bibitem[{{Shen} \& {Bildsten}(2009)}]{shen09}
{Shen} K.~J., {Bildsten} L., 2009, APJ, 699, 1365

\bibitem[{{Shen} \& {Bildsten}(2014)}]{shen14a}
{Shen} K.~J., {Bildsten} L., 2014, ApJ, 785, 61

\bibitem[{{Shen} {et~al}\mbox{.}(2010){Shen}, {Kasen}, {Weinberg}, {Bildsten},
  \& {Scannapieco}}]{shen10}
{Shen} K.~J., {Kasen} D., {Weinberg} N.~N., {Bildsten} L., {Scannapieco} E.,
  2010, ApJ, 715, 767

\bibitem[{{Shen} \& {Moore}(2014)}]{shen14}
{Shen} K.~J., {Moore} K., 2014, ApJ, 797, 46

\bibitem[{{Sim} {et~al}\mbox{.}(2012){Sim}, {Fink}, {Kromer}, {R{\"o}pke},
  {Ruiter}, \& {}}]{sim12}
{Sim} S.~A., {Fink} M., {Kromer} M., {R{\"o}pke} F.~K., {Ruiter} A.~J., {} W.,
  2012, MNRAS, 420, 3003

\bibitem[{{Sim} {et~al}\mbox{.}(2010){Sim}, {R{\"o}pke}, {Hillebrandt},
  {Kromer}, {Pakmor}, {Fink}, {Ruiter}, \& {Seitenzahl}}]{sim10}
{Sim} S.~A., {R{\"o}pke} F.~K., {Hillebrandt} W., {Kromer} M., {Pakmor} R.,
  {Fink} M., {Ruiter} A.~J., {Seitenzahl} I.~R., 2010, ApJL, 714, L52

\bibitem[{{Stehle} {et~al}\mbox{.}(2005){Stehle}, {Mazzali}, {Benetti}, \&
  {Hillebrandt}}]{stehle05}
{Stehle} M., {Mazzali} P.~A., {Benetti} S., {Hillebrandt} W., 2005, MNRAS, 360,
  1231

\bibitem[{{Stritzinger} {et~al}\mbox{.}(2006){Stritzinger}, {Mazzali},
  {Sollerman}, \& {Benetti}}]{stritzinger06}
{Stritzinger} M., {Mazzali} P.~A., {Sollerman} J., {Benetti} S., 2006, A\&A,
  460, 793

\bibitem[{{Tanaka} {et~al}\mbox{.}(2011){Tanaka}, {Mazzali}, {Stanishev},
  {Maurer}, {Kerzendorf}, \& {Nomoto}}]{tanaka11}
{Tanaka} M., {Mazzali} P.~A., {Stanishev} V., {Maurer} I., {Kerzendorf} W.~E.,
  {Nomoto} K., 2011, MNRAS, 410, 1725

\bibitem[{{Timmes}(1999)}]{timmes99}
{Timmes} F.~X., 1999, ApJS, 124, 241

\bibitem[{{Timmes} \& {Swesty}(2000)}]{timmes00}
{Timmes} F.~X., {Swesty} F.~D., 2000, ApJS, 126, 501

\bibitem[{{Toonen} {et~al}\mbox{.}(2012){Toonen}, {Nelemans}, \& {Portegies
  Zwart}}]{toonen12}
{Toonen} S., {Nelemans} G., {Portegies Zwart} S., 2012, A\&A, 546, A70

\bibitem[{{Tsebrenko} \& {Soker}(2015)}]{tsebrenko15}
{Tsebrenko} D., {Soker} N., 2015, MNRAS, 447, 2568

\bibitem[{{Waldman} {et~al}\mbox{.}(2011){Waldman}, {Sauer}, {Livne}, {Perets},
  {Glasner}, {Mazzali}, {Truran}, \& {Gal-Yam}}]{waldman11}
{Waldman} R., {Sauer} D., {Livne} E., {Perets} H., {Glasner} A., {Mazzali} P.,
  {Truran} J.~W., {Gal-Yam} A., 2011, ApJ, 738, 21

\bibitem[{{Wang} \& {Han}(2012)}]{wang12}
{Wang} B., {Han} Z., 2012, NewAR, 56, 122

\bibitem[{{Webbink}(1984)}]{webbink84}
{Webbink} R.~F., 1984, ApJ, 277, 355

\bibitem[{{Whelan} \& {Iben}(1973)}]{whelan73}
{Whelan} J., {Iben}, Jr. I., 1973, ApJ, 186, 1007

\bibitem[{{Woosley} \& {Kasen}(2011)}]{woosley11}
{Woosley} S.~E., {Kasen} D., 2011, ApJ, 734, 38

\bibitem[{{Woosley} \& {Weaver}(1994)}]{woosley94}
{Woosley} S.~E., {Weaver} T.~A., 1994, ApJ, 423, 371

\bibitem[{{Yamaguchi} {et~al}\mbox{.}(2014){Yamaguchi}, {Badenes}, {Petre},
  {Nakano}, {Castro}, {Enoto}, {Hiraga}, {Hughes}, {Maeda}, {Nobukawa},
  {Safi-Harb}, {Slane}, {Smith}, \& {Uchida}}]{yamaguchi14}
{Yamaguchi} H. {et~al.}, 2014, ApJL, 785, L27

\bibitem[{{Yoon} {et~al}\mbox{.}(2007){Yoon}, {Podsiadlowski}, \&
  {Rosswog}}]{yoon07}
{Yoon} S., {Podsiadlowski} P., {Rosswog} S., 2007, MNRAS, 380, 933

\bibitem[{{Zhu} {et~al}\mbox{.}(2013){Zhu}, {Chang}, {van Kerkwijk}, \&
  {Wadsley}}]{zhu13}
{Zhu} C., {Chang} P., {van Kerkwijk} M.~H., {Wadsley} J., 2013, ApJ, 767, 164

\bibitem[{{Zoglauer} {et~al}\mbox{.}(2015){Zoglauer}, {Reynolds}, {An},
  {Boggs}, {Christensen}, {Craig}, {Fryer}, {Grefenstette}, {Harrison},
  {Hailey}, {Krivonos}, {Madsen}, {Miyasaka}, {Stern}, \& {Zhang}}]{zoglauer15}
{Zoglauer} A. {et~al.}, 2015, ApJ, 798, 98

\end{thebibliography}


\appendix

\setcounter{table}{0} \renewcommand{\thetable}{A\arabic{table}}

\subsection{Initial conditions}
\label{sec:App.ICs}

\begin{small}
\begin{table}
\begin{tabular}{|c|c|c|c|c|}
  \hline\hline
  Run & Initial & Initial & \multirow{2}{*}{$T_{\rm max,8}$} & \multirow{2}{*}{$\rho_5(T_{\rm max})$} \\
   No & masses [$M_\odot$] & compositions & &  \\[0.5ex]
  \hline\hline
 \multicolumn{5}{|c|}{Helium mass transferring systems}\\[0.5ex]
  \hline
  1 & $0.45+0.45$ & He-He & 0.990 & 0.543 \\
  \hline
  2 & $0.45+0.6$ & He-HeCO & 1.934 & 0.436 \\
  \hline
  3 & $0.45+0.9$ & He-CO & 4.799 & 0.201 \\
  \hline
  4 & $0.45+1.1$ & He-ONe & 5.235 & 3.872\\
  \hline
  5 &$0.55+1$ & HeCO-CO & 5.384 & 3.941\\
  \hline
  6 & $0.6+0.6$ & HeCO-HeCO & 3.341 & 0.720 \\
  \hline
  7 & $0.6+1.1$ & HeCO-CO & 7.218 & 6.354\\
  \hline
\multicolumn{5}{|c|}{Carbon/oxygen mass transferring systems}\\[0.5ex]
  \hline
  8 & $0.95+1.15$ &CO-ONe & 9.693 & 7.411\\
    \hline
  9 & $1.05+1.05$ & CO-CO & 7.844 & 19.523\\
\hline\hline
\end{tabular}
\caption{Initial conditions of the first set of runs, without perturbations:
$T_{\rm max,8}$ (in units of $10^8$ K) and $\rho_5(T_{\rm max})$ (in units of $10^5\, {\rm g/cm^{-3}}$) 
are the maximum temperature and density at the location of maximum temperature, respectively. The masses 
(in $M_\odot$) and compositions are those from the beginning of the SPH calculations.}
\label{tab:ICsnoper}
\end{table}
\end{small}
\begin{center}
\begin{small}
\begin{table*}
\begin{tabular}{|c|c|c|c|c|c|c|c|c|}
  \hline\hline
  Run & 
        Initial & Initial & \multirow{2}{*}{$s_{\rm per,9}$} & \multirow{2}{*}{$z_{\rm
                                                               per,9}$} &\multirow{2}{*}{$T_{\rm per,8}$} &
                                                                                                            \multirow{2}{*}{$\rho_5(T_{\rm
                                                                                                            per})$} &
    hotspot & \multirow{2}{*}{Comments} \\
    No & masses [$M_\odot$] & compositions & & 
       & & & criteria & \\[0.5ex]
  \hline\hline
  \multicolumn{9}{|c|}{Helium mass transferring systems}\\[0.5ex]
    \hline
    1.c1 & $0.45+0.45 $ & He-He & 0.403 & 0.661 & 1.01 & 0.776  & $C_1$ & quadrant I
    \\
    \hline
    1.c2 & $0.45+0.45 $ & He-He & 0.460 & 0.882 & 1.292 & 0.100  & $C_2$ & quadrant I
    \\
    \hline
    1.c3 & $0.45+0.45 $ & He-He & 0.247 & 0.653 & 1.219 & 1.107  & $C_3$ & quadrant I\\
    \hline
    2.c1 & $0.45+0.6 $ & He-HeCO & 0.744 & 0.162 & 1.980 & 1.976  & $C_1$&
                                                                   quadrant IV
    \\
    \hline
    2.c2 & $0.45+0.6 $ & He-HeCO & 0.532 & 0.773 & 2.336 & 0.196  & $C_2$ & quadrant I
    \\
    \hline
    2.c3 & $0.45+0.6 $ & He-HeCO & 0.703 & 0.389 & 2.092 & 1.085  & $C_3$ & quadrant I
    \\
    \hline
    3.c1/c3& \multirow{1}{*}{$0.45+0.9 $} & \multirow{1}{*}{He-CO} & \multirow{1}{*}{0.681} & \multirow{1}{*}{0.029} & \multirow{1}{*}{5.525} & \multirow{1}{*}{1.229}  & $C_1,\,C_3$ & \multirow{1}{*}{quadrant I}
    \\
    \hline
     3.c2& $0.45+0.9 $ & He-CO & 0.395 & 0.645 & 5.948 & 0.313  & $C_2$ & quadrant I
    \\
    \hline
  4.c1/c2 & \multirow{1}{*}{$0.45+1.1 $} &
                                                         \multirow{1}{*}{He-ONe}
                            & \multirow{1}{*}{0.595} &
                                                       \multirow{1}{*}{0.026}  & \multirow{1}{*}{6.810} & \multirow{1}{*}{0.895} & $C_1$ & \multirow{1}{*}{quadrant I}
    \\
    \hline
    
    \multirow{1}{*}{4.c3}& \multirow{1}{*}{$0.45+1.1 $} & \multirow{1}{*}{He-ONe} & \multirow{1}{*}{0.575} & \multirow{1}{*}{0.034} & \multirow{1}{*}{4.698} & \multirow{1}{*}{1.166}  &
                                                                     \multirow{1}{*}{$C_3$}
             & no run, as $T_{\rm per}\approx T_{\rm grid}$
    \\
    \hline
    5.c1/c2/c3& \multirow{1}{*}{$0.55+1 $} &
                                                        \multirow{1}{*}{HeCO-CO} & \multirow{1}{*}{0.607} & \multirow{1}{*}{0.009} & \multirow{1}{*}{9.790} & \multirow{1}{*}{2.320}  & $C_1,\,C_2,\,C_3$&
                                                                    \multirow{1}{*}{quadrant IV}\\
     \hline
    6.c1/c2/c3& \multirow{1}{*}{$0.6+0.6 $} & \multirow{1}{*}{HeCO-HeCO} & \multirow{1}{*}{0.725} & \multirow{1}{*}{0.307} & \multirow{1}{*}{3.670} & \multirow{1}{*}{1.629}  & $C_1,\,C_2,\,C_3$ &
    \multirow{1}{*}{quadrant IV}\\
     \hline
    7.c1/c3& \multirow{1}{*}{$0.6+1.1 $} & \multirow{1}{*}{HeCO-CO} &
                                                                   \multirow{1}{*}{0.640} & \multirow{1}{*}{0.085} & \multirow{1}{*}{4.640} & \multirow{1}{*}{1.282}  & $C_1,\,C_3$&
                                                                     \multirow{1}{*}{quadrant IV}
    \\
\hline
    7.c2& $0.6+1.1 $ & HeCO-CO & 0.818 & 0.131 & 6.243 & 0.544  & $C_2$& quadrant I
  \\[0.5ex]
    \hline
    \multicolumn{9}{|c|}{Carbon/oxygen mass transferring systems}\\[0.5ex]
    \hline
    8.c1 & $0.95+1.15 $ & CO - ONe & 0.295 & 0.173 & 11.6 & 28.525  & 
                                                              
                                                                     $C_1$
             & quandrant IV\\
    \hline
    8.c2/c4 & \multirow{1}{*}{$0.95+1.15 $} & \multirow{1}{*}{CO - ONe} & \multirow{1}{*}{0.278} & \multirow{1}{*}{0.248} & \multirow{1}{*}{11.738} & \multirow{1}{*}{14.898}  & $C_2,\, C_4$ & \multirow{1}{*}{quadrant I}\\
  \hline
  9.c1/c2/c4 & \multirow{1}{*}{$1.05+1.05$}  & \multirow{1}{*}{CO-CO} & \multirow{1}{*}{0.405} & \multirow{1}{*}{0.029} & \multirow{1}{*}{10.407} & \multirow{1}{*}{34.661}  & $C_1,\, C_2,\, C_4$ &
                                                                     \multirow{1}{*}{quadrant IV}\\                                     
\hline\hline
\end{tabular}
\caption{Initial conditions for the second set of runs, with perturbations based 
on different criteria guided by the 3D SPH simulations. 
  $s_{\rm per,9}$ and $z_{\rm per,9}$ are the coordinates on the $s$-- and
    $z$--axis where the hotspot's center is placed, respectively,
  $T_{\rm per,8}$ is the perturbation temperature (in units of
    $10^8\, {\rm K}$) and $\rho_5(T_{\rm per})$ is the density at the
  location of the perturbation (in units of
  $10^5\, {\rm g\, cm^{-3}}$). The hotspot criteria labelling is
  explained in Section \ref{sec:ICs}. }  
  \label{tab:ICs}
\end{table*}
\end{small}
\end{center}

We run three sets of simulations, with and without a temperature perturbation. 
Table \ref{tab:ICsnoper} shows the initial conditions for the first set of runs,
without temperature perturbations. The initial conditions for the second set of 
runs, where temperature perturbations are set based  
on different criteria guided by the 3D SPH simulations,
are given in Table \ref{tab:ICs}.
For the third set of runs, the temperature perturbations are setup following 
the critical conditions for a spontaneous initiation of a detonation from the 
spatially resolved 1D calculations of \cite{holcomb13} and \cite{shen14} for 
He composition and \cite{roepke07} and \cite{seitenzahl09} for CO and 
the initial conditions are given in Table \ref{tab:ICsdd}. 

Runs are ordered by increasing donor's mass and by decreasing mass ratio 
$q$ and labeled 1 through 9. For the second and third set of runs, 
we have added to the run number the hotspot search criteria and a ``.dd'', 
respectively.

\begin{center}
\begin{small}
\begin{table*}
\begin{tabular}{|c|c|c|c|c|c|c|c|}
  \hline\hline
  Run & Initial & Initial & \multirow{2}{*}{$s_{\rm per,9}$} & \multirow{2}{*}{$z_{\rm per,9}$} &\multirow{2}{*}{$T_{\rm per,8}$} &
                                                                                                            \multirow{2}{*}{$\rho_5(T_{\rm per})$} &
    \multirow{2}{*}{Comments} \\
    No & masses [$M_\odot$] & compositions & & 
       & &  & \\[0.5ex]
  \hline
  \multicolumn{8}{|c|}{Helium mass transferring systems}\\[0.5ex]
  \hline
   \multirow{2}{*}{1.dd} & \multirow{2}{*}{$0.45+0.45$} & \multirow{2}{*}{He-He} & \multirow{2}{*}{0.546} & \multirow{2}{*}{0.184} & \multirow{2}{*}{5.0} & \multirow{2}{*}{5.012}  & unrealistic ICs: hotspot inside \\
  & & & & & & & cold (dense) core\\
  \hline
  2.dd & $0.45+0.6$ & He-HeCO & 0.576 & 0.166 & 5.0 & 5.002  & \\
  \hline
  2.dd-Polar & $0.45+0.6$ & He-HeCO & 0.0 & 0.522 & 5.0 & 5.045  & \\
  \hline
  3.dd & $0.45+0.9$ & He-CO & 0.584 & 0.071 & 8.0 & 3.008  & \\
  \hline
  3.dd-Polar & $0.45+0.9$ & He-CO & 0.0 & 0.593 & 8.0 & 1.713  & \\
  \hline
  \multirow{2}{*}{5.dd} & \multirow{2}{*}{$0.55+1.0$} & \multirow{2}{*}{HeCO-CO} & \multirow{2}{*}{0.322} & \multirow{2}{*}{0.369} & \multirow{2}{*}{8.0} & \multirow{2}{*}{5.011}  & hotspot composition set to $X(\!\,^4{\rm He})=1$, \\
& & & & & & & as $X(\!\,^4{\rm He})<0.1$ at $\rho\geq 6\times 10^4\,{\rm g\, cm^{-3}}$\\
\hline
  6.dd & $0.6+0.6$ & HeCO-HeCO & 0.225 & 0.539 & 5.0 & 3.011 & \\
\hline
  \multirow{2}{*}{7.dd} & \multirow{2}{*}{$0.6+1.1$} & \multirow{2}{*}{HeCO-ONe} & \multirow{2}{*}{0.387} & \multirow{2}{*}{0.238} & \multirow{2}{*}{-} & \multirow{2}{*}{5.966} & No perturb., but pure He at ($s_{\rm per},z_{\rm per}$),\\
& & & & & & & as $X(\!\,^4{\rm He})<0.1$ at $\rho\geq 3.9\times 10^4\,{\rm g\, cm^{-3}}$\\
\hline
\multicolumn{8}{|c|}{Carbon/oxygen mass transferring systems}\\[0.5ex]
  \hline
8.dd & $0.95+1.15$ & CO-ONe & 0.341 & 0.083 & 30.0 & 30.313 & \\
\hline
9.dd & $1.05+1.05$ & CO-CO & 0.135 & 0.308 & 30.0 & 19.923 & \\
\hline\hline
\end{tabular}
\caption[Initial conditions for the third set of runs, with perturbations based on the spatially 
resolved 1D calculations]{Initial conditions for the third set of runs, using the critical conditions for a spontaneous 
  initiation of a detonation from the 
  spatially resolved 1D calculations of \cite{holcomb13} and \cite{shen14} for He composition and
\cite{roepke07} and \cite{seitenzahl09} for CO.  Same quantities as in Table \ref{tab:ICs},
  only labelling has been changed, with ``dd'' added after the run number. } 
  \label{tab:ICsdd}
\end{table*}
\end{small}
\end{center}

\subsection{Nucleosynthetic yields}
\label{sec:App.yields}

The nucleosynthetic yields and kinetic energies for all the models that detonated. Run 4 is the only run 
without a perturbation where 
a single, He-detonation is triggered, while for run 4.c1/c2, with a
perturbation of $R_{\rm per}=500\, {\rm km}$ at the location where
the ratio $\tau_{\rm nuc}/\tau_{\rm dyn}(T)$ is minimum in the SPH calculations, a double
detonation is triggered. 
Runs labelled with ``dd'' after the number are those where the hotspots are setup 
following the critical conditions for a direct initiation of a detonation from the
spatially resolved 1D calculations. For this set of runs, for all models but 7.dd a detonation 
is triggered in the core.

\begin{center}
\begin{small}
\begin{table*}
\setlength{\tabcolsep}{5pt}
  \begin{tabular}{|c|c|c|c|c|c|c|c|c|c|c|c|c|c|c|}
    \hline\hline
& 1.dd & 2.dd & 3.dd & 4 & 4.c1/c2 & 5.dd  & 6.dd & 7.dd & 8.dd &
                                                                  9.dd\\
& $0.45\!+\!0.45$ & $0.45\!+\!0.6$ & $0.45\!+\!0.9$
                     & $0.45\!+\!1.1$ & $0.45\!+\!1.1$ & $0.55\!+\!1$  & $0.6\!+\!0.6$ & $0.6\!+\!1.1$ & $0.95\!+\!1.15$ &
                                                                  $1.05\!+\!1.05$\\
\hline
$\!\,^{4}{\rm He}$ & $4.95(-1)$ & $4.03(-1)$ & $4.14(-1)$   & $3.67(-1)$ & $4.47(-2)$ & $8.73(-2)$ & $8.05(-2)$ & $9.45(-2)$ & $8.55(-3)$ & $4.34(-3)$\\[2pt]
$\!\,^{12}{\rm C}$ & $9.83(-3)$& $1.54(-2)$ &  $7.03(-2)$    & $2.77(-2)$ & $7.13(-3)$ & $1.91(-1)$ & $2.35(-1)$ & $2.27(-1)$ & $2.04(-1)$ & $1.94(-1)$\\[2pt]
$\!\,^{16}{\rm O}$ & $1.33(-5)$ & $1.57(-1)$ & $2.13(-1)$    & $6.39(-1)$ & $1.83(-1)$ & $2.76(-1)$& $2.54(-1)$ & $8.27(-1)$  & $5.95(-1)$ & $6.21(-1)$\\[2pt]
$\!\,^{20}{\rm Ne}$ & $3.44(-5)$ & $2.11(-3)$ & $2.08(-2)$   & $3.84(-1)$ & $8.95(-2)$ & $6.14(-2)$& $7.30(-2)$ & $3.94(-1)$ & $4.75(-2)$ & $2.72(-2)$\\[2pt] 
$\!\,^{24}{\rm Mg}$ &  $8.08(-5)$ & $6.35(-2)$ & $3.53(-2)$ & $6.34(-2)$ & $2.65(-2)$ &  $4.62(-2)$ & $1.56(-1)$ & $6.28(-2)$ & $4.38(-2)$ & $6.22(-2)$\\[2pt] 
$\!\,^{28}{\rm Si}$ & $2.3(-3)$ & $1.94(-1)$ & $3.04(-1)$      & $1.02(-2)$ & $1.86(-1)$ & $3.17(-1)$ & $3.21(-1)$ & $5.87(-3)$ & $2.29(-1)$ & $3.58(-1)$\\[2pt]
$\!\,^{32}{\rm S}$ & $3.2(-3)$ & $4.46(-2)$ & $1.5(-1)$         & $3.52(-3)$ & $1.13(-1)$ & $1.63(-1)$  & $4.79(-2)$ & $7.31(-4)$ & $1.09(1)$ & $1.64(-1)$\\[2pt]
 $\!\,^{36}{\rm Ar}$ & $7.18(-3)$ & $2.29(-2)$ & $2.89(-2)$  & $1.01(-3)$ & $2.36(-2)$ & $3.14(-2)$ & $9.45(-3)$ & $5.63(-4)$  & $2.17(-2)$ & $3.07(-2)$\\[2pt]
$\!\,^{40}{\rm Ca}$ & $3.25(-2)$ & $5.48(-2)$ & $2.35(-2)$   & $1.55(-4)$ & $2.35(-2)$ & $2.97(-2)$ & $4.69(-3)$ & $2.08(-3)$ & $2.12(-2)$ & $2.82(-2)$\\[2pt]
$\!\,^{44}{\rm Ti}$ & $6.14(-2)$ & $4.66(-2)$ & $2.53(-5)$    & $2.02(-6)$ & $2.21(-5)$ & $2.21(-3)$ & $4.13(-3)$ & $2.31(-3)$ & $4.01(-5)$ & $2.91(-5)$\\[2pt]
$\!\,^{48}{\rm Cr}$ & $3.8(-2)$ & $2.22(-2)$ & $1.80(-4)$   & $2.13(-9)$ & $3.46(-4)$ & $2.23(-3)$  & $3.62(-3)$ & $1.91(-3)$ & $3.03(-4)$ & $3.12(-4)$\\[2pt]
$\!\,^{52}{\rm Fe}$ & $3.5(-2)$ & $9.27(-3)$ & $3.12(-3)$     & $4.68(-13)$ & $8.29(-3)$ & $9.95(-3)$& $1.62(-3)$ & $4.93(-3)$ & $6.65(-3)$  & $6.59(-3)$\\[2pt]
$\!\,^{56}{\rm Ni}$ & $2.21(-1)$ & $1.25(-2)$ & $8.76(-2)$   & $3.37(-13)$ & $4.41(-1)$ & $3.01(-1)$ & $6.81(-4)$ & $3.48(-2)$ & $8.22(-1)$  & $5.41(-1)$\\[2pt]
\hline
$28\!\leq\! A\!\leq\! 40$ & 0.045 & 0.158 & 0.506  & 0.015 & 0.346 & 0.54 & 0.383 & 0.009 & 0.38 & 0.581 \\
$A\geq 44$             & 0.355 & 0.045  & 0.091 & 2.019(-6) & 0.45  & 0.315 & 0.01  & 0.044 & 0.829 & 0.548 \\
\hline
$E_{\rm kin}$ & $1.09(51)$ & $7.54(50)$ & $6.76(50)$ &
                                                                     $1.75(49)$
                         & $6.87(50)$ & $1.05(51)$ & $7.46(50)$ &
                                                                  $9.08(49)$ & $7.149(50)$ & $1.27(51)$\\
\hline
$\langle v_{\rm Si}\rangle $ & 11.2  & 5.4 & 5.6 & 1.4 & 8.4 & 8.2 & 6.8 & 5.5 & 8.1 & 7.4 \\
$\langle v_{\rm Ca}\rangle $ & 11.5  & 9.9 & 4.9 & 1.5 & 7.5 & 7.6 & 8.4 & 6.8 & 7.6 & 6.8 \\
\hline\hline 
\end{tabular}
\caption{Isotope yields (by mass, in units of $M_\odot$), kinetic
  energies (in units of $\rm {erg}$) and the Si and Ca mass-weighted
  velocities (in units of $10^3\, {\rm km/s}$)  for all runs where
  detonations are triggered. Below the run numbers are the 
  initial masses (in $M_\odot$) of the two stars. The floating-point
  numbers are expressed in the form {\em significand} 
followed in the brackets by the {\em exponent} of the base 10.}
\label{tab:nuc}
\end{table*}
\end{small}
\end{center}

\label{lastpage}

\end{document}